\pdfoutput=0
\documentclass[11pt]{article}
\usepackage{axodraw}
\usepackage{epsfig}
\usepackage{amsfonts}
\usepackage{amsmath}
\usepackage{bm,bbm}
\usepackage{cite}
\allowdisplaybreaks[1]

\input pix.sty

\newcommand{\la}[1]{\label{#1}}
\newcommand{\be}{\begin{equation}}
\newcommand{\ee}{\end{equation}}
\newcommand{\ba}{\begin{eqnarray}}
\newcommand{\ea}{\end{eqnarray}}
\newcommand{\rmi}[1]{{\mbox{\scriptsize #1}}}
\newcommand{\fig}{Fig.~}

\newcommand{\eq}{Eq.~}
\newcommand{\eqs}{Eqs.~}
\newcommand{\se}{Sec.~}

\newcommand{\nr}[1]{(\ref{#1})}
\newcommand{\tr}{{\rm Tr\,}}
\newcommand{\nn}{\nonumber \\}
\newcommand{\fr}[2]{{\frac{#1}{#2}\,}}
\newcommand{\msbar}{{\overline{\mbox{\rm MS}}}}

\renewcommand{\vec}[1]{{\bf #1}}

\renewcommand{\eq}{eq.~}
\renewcommand{\eqs}{eqs.~}
\renewcommand{\se}{sec.~}

\renewcommand{\fig}{fig.~}

\newcommand{\aL}{a^{ }_\rmii{L}}
\newcommand{\aR}{a^{ }_\rmii{R}}

\newcommand{\gammaE}{\gamma_\rmii{E}}
\newcommand{\rmO}{{\mathcal{O}}}
\newcommand{\bmu}{\bar\mu}

\newcommand{\CF}{C_\rmii{F}}

\def\lsi{\raise0.3ex\hbox{$<$\kern-0.75em\raise-1.1ex\hbox{$\sim$}}}
\def\gsi{\raise0.3ex\hbox{$>$\kern-0.75em\raise-1.1ex\hbox{$\sim$}}}

\newcommand{\gsim}{\mathop{\gsi}}

\newcommand{\sign}{\mathop{\mbox{sign}}}
\newcommand{\nF}{n_\rmii{F}}
\newcommand{\nB}{n_\rmii{B}}
\newcommand{\rmii}[1]{{\mbox{\tiny\rm{#1}}}}
\newcommand{\rmiii}[1]{{\mbox{\tiny{$\scriptstyle{\rm#1}$}}}}
\newcommand{\re}{\mathop{\mbox{Re}}}
\newcommand{\im}{\mathop{\mbox{Im}}}
\newcommand{\Tint}[1]{{\hbox{$\sum$}\!\!\!\!\!\!\!\int\,}_{\!\!\!\!\raise-0.9ex\hbox{$\scriptstyle{#1}$}}}
\newcommand{\Tinti}[1]{{{\Sigma}\!\!\!\!\raise0.3ex\hbox{$\int$}_\rmii{${#1}$}}}
\newcommand{\bi}{\begin{itemize}}
\newcommand{\ei}{\end{itemize}}
\newcommand{\hide}[1]{ }
\newcommand{\bsl}[1]{\,\slash\!\!\!\!{#1}\,}
\newcommand{\msl}[1]{\,\slash\!\!\!{#1}\,}
\newcommand{\deltabar}{\raise-0.02em\hbox{$\bar{}$}\hspace*{-0.8mm}{\delta}}
\newcommand{\ddeltabar}{\raise-0.18em\hbox{$\bar{}$}\hspace*{-0.8mm}{\delta}}
\renewcommand{\P}{\mathcal{P}}

\newcommand{\K}{\mathcal{K}}

\newcommand{\E}{\rmii{$E$}}

\newcommand{\M}{\rmii{$M$}}
\newcommand{\T}{\rmii{$T$}}

\newcommand{\aW}{\rmii{$W$}}
\newcommand{\xW}{s^2_\aW}

\newcommand{\mW}{m_\rmii{$W$}} 
\newcommand{\lnf}{l^{ }_\rmi{1f}}
\newcommand{\lif}{l^{ }_\rmi{2f}}
\newcommand{\ltf}{l^{ }_\rmi{3f}}
\def\TAsc(#1,#2)(#3,#4,#5)%
{\SetWidth{2.0}\CArc(#1,#2)(#3,#4,#5)\SetWidth{1.0}}
\def\TLsc(#1,#2)(#3,#4)%
{\SetWidth{2.0}\Line(#1,#2)(#3,#4)\SetWidth{1.0}}
\def\Lwidth{3}

\def\TAgl(#1,#2)(#3,#4,#5){\SetWidth{2.0}\PhotonArc(#1,#2)(#3,#4,#5){\Lwidth}%
{6.283 #3 mul 360 div #4 #5 sub #4 #5 sub mul sqrt mul Tdensity mul}%
\SetWidth{1.0}}
\def\TLgl(#1,#2)(#3,#4){\SetWidth{2.0}\Photon(#1,#2)(#3,#4){\Lwidth}
{#1 #3 sub #1 #3 sub mul #2 #4 sub #2 #4 sub mul add sqrt Tdensity mul}%
\SetWidth{1.0}}

\renewcommand{\picc}[1]{\;\parbox[c]{60pt}{\begin{picture}(60,30)(0,-10)
\SetWidth{1.0}\SetScale{1.3} #1 \end{picture}}\; }
\def\Lwidth{1.3}

\def\cutNLO{\picb{%
 \Lqu(0,0)(15,10)%
 \Lqu(35,10)(50,0)%
 \Aaqu(25,20)(13,90,230)%
 \CArc(25,20)(13,-50,90)%
 \CCirc(15,10){3}{Black}{Black}%
 \CCirc(35,10){3}{Black}{Black}%
 \CCirc(25,10){10}{Black}{Gray}%
 \Line(25,-8)(25,38)
 \Line(20,-10)(25,-8)
 \Line(25,38)(30,40)
}}
\def\NLOblob{\picb{%
 \CCirc(15,15){3}{Black}{Black}%
 \CCirc(35,15){3}{Black}{Black}%
 \CCirc(25,15){10}{Black}{Gray}%
}}
\def\selfEa{\picb{%
 \TAsc(25,15)(15,0,360)%
 \Lgl(25,0)(25,30)
 \CCirc(10,15){3}{Black}{Black}%
 \CCirc(40,15){3}{Black}{Black}%
}}
\def\selfEb{\picb{%
 \TAsc(25,15)(15,0,360)%
 \Agl(25,30)(10,200,340)
 \CCirc(10,15){3}{Black}{Black}%
 \CCirc(40,15){3}{Black}{Black}%
}}
\def\selfEc{\picb{%
 \TAsc(25,15)(15,0,360)%
 \Agl(25,0)(10,20,160)
 \CCirc(10,15){3}{Black}{Black}%
 \CCirc(40,15){3}{Black}{Black}%
}}
\def\selfEd{\picb{%
 \TLsc(8,11)(16,11)%
 \TLsc(8,11)(8,19)%
 \TLsc(16,11)(16,19)%
 \TLsc(8,19)(16,19)%
 \Lgl(16,15)(34,15)%
 \Text(17,15)[c]{$*$}
 \TLsc(42,11)(34,11)%
 \TLsc(42,11)(42,19)%
 \TLsc(34,11)(34,19)%
 \TLsc(42,19)(34,19)%
}}
\def\selfEf{\picb{%
 \TLsc(8,11)(16,11)%
 \TLsc(8,11)(8,19)%
 \TLsc(16,11)(16,19)%
 \TLsc(8,19)(16,19)%
 \Lgl(16,15)(25,15)%
 \Text(23.5,10.2)[c]{$\equiv$}%
 \TAsc(56,15)(10,0,360)%
 \CCirc(46,15){3}{Black}{Black}%
 \Lgl(66,15)(75,15)%
 \Text(58,10.2)[c]{$+$}%
 \CBoxc(94,15)(5,5){Black}{Black}%
 \Lgl(96,15)(105,15)%
}}
\def\selfEfX{\picb{%
 \Lqu(-2,0)(6,14)%
 \Laqu(-2,30)(6,16)%
 \TLsc(6,11)(14,11)%
 \TLsc(6,11)(6,19)%
 \TLsc(14,11)(14,19)%
 \TLsc(6,19)(14,19)%
 \Lgl(14,15)(23,15)%
 \Text(22.5,10.2)[c]{$\equiv$}%
 \Lqu(40,0)(48,14)%
 \Laqu(40,30)(48,16)%
 \TAsc(58,15)(10,0,360)%
 \CCirc(48,15){3}{Black}{Black}%
 \Lgl(68,15)(77,15)%
 \Text(60,10.2)[c]{$+$}%
 \Lqu(88,0)(96,14)%
 \Laqu(88,30)(96,16)%
 \CBoxc(98,15)(5,5){Black}{Black}%
 \Lgl(100,15)(109,15)%
}}
\def\ampLOsa{\picb{%
 \Lqu(0,0)(25,15)%
 \Lqu(25,15)(0,30)%
 \CCirc(25,15){3}{Black}{Black}%
 \TLsc(50,0)(25,15)%
 \TLsc(25,15)(50,30)%
}}
\def\ampLOsb{\picb{%
 \Lqu(50,30)(25,15)%
 \Lqu(25,15)(50,0)%
 \CCirc(25,15){3}{Black}{Black}%
 \TLsc(0,30)(25,15)%
 \TLsc(25,15)(0,0)%
}}
\def\ampLOt{\picb{%
 \Lqu(0,0)(25,15)%
 \Lqu(25,15)(50,0)%
 \CCirc(25,15){3}{Black}{Black}%
 \TLsc(0,30)(25,15)%
 \TLsc(25,15)(50,30)%
}}
\def\virtNLOsa{\picb{%
 \Lqu(0,0)(25,15)%
 \Lqu(25,15)(0,30)%
 \Agl(25,15)(15,-30,30)
 \CCirc(25,15){3}{Black}{Black}%
 \TLsc(50,0)(25,15)%
 \TLsc(25,15)(50,30)%
}}
\def\virtNLOsb{\picb{%
 \Lqu(0,0)(25,15)%
 \Lqu(25,15)(0,30)%
 \Agl(38,23.4)(8,210,30)
 \CCirc(25,15){3}{Black}{Black}%
 \TLsc(50,0)(25,15)%
 \TLsc(25,15)(50,30)%
}}
\def\virtNLOsc{\picb{%
 \Lqu(0,0)(25,15)%
 \Lqu(25,15)(0,30)%
 \Agl(38,6.6)(8,-30,150)
 \CCirc(25,15){3}{Black}{Black}%
 \TLsc(50,0)(25,15)%
 \TLsc(25,15)(50,30)%
}}
\def\virtNLOsd{\picb{%
 \Lqu(50,30)(25,15)%
 \Lqu(25,15)(50,0)%
 \Agl(25,15)(15,150,210)
 \CCirc(25,15){3}{Black}{Black}%
 \TLsc(0,30)(25,15)%
 \TLsc(25,15)(0,0)%
}}
\def\virtNLOse{\picb{%
 \Lqu(50,30)(25,15)%
 \Lqu(25,15)(50,0)%
 \Agl(12,23.4)(8,150,-30)
 \CCirc(25,15){3}{Black}{Black}%
 \TLsc(0,30)(25,15)%
 \TLsc(25,15)(0,0)%
}}
\def\virtNLOsf{\picb{%
 \Lqu(50,30)(25,15)%
 \Lqu(25,15)(50,0)%
 \Agl(12,6.6)(8,30,210)
 \CCirc(25,15){3}{Black}{Black}%
 \TLsc(0,30)(25,15)%
 \TLsc(25,15)(0,0)%
}}
\def\virtNLOta{\picb{%
 \Lqu(0,0)(25,15)%
 \Lqu(25,15)(50,0)%
 \Agl(25,15)(15,30,150)
 \CCirc(25,15){3}{Black}{Black}%
 \TLsc(0,30)(25,15)%
 \TLsc(25,15)(50,30)%
}}
\def\virtNLOtb{\picb{%
 \Lqu(0,0)(25,15)%
 \Lqu(25,15)(50,0)%
 \Agl(12,23.4)(8,-30,150)
 \CCirc(25,15){3}{Black}{Black}%
 \TLsc(0,30)(25,15)%
 \TLsc(25,15)(50,30)%
}}
\def\virtNLOtc{\picb{%
 \Lqu(0,0)(25,15)%
 \Lqu(25,15)(50,0)%
 \Agl(38,23.4)(8,30,210)
 \CCirc(25,15){3}{Black}{Black}%
 \TLsc(0,30)(25,15)%
 \TLsc(25,15)(50,30)%
}}
\def\realNLOsa{\picb{%
 \Lqu(0,0)(25,15)%
 \Lqu(25,15)(0,30)%
 \Lgl(38,23.4)(50,15)
 \CCirc(25,15){3}{Black}{Black}%
 \TLsc(50,0)(25,15)%
 \TLsc(25,15)(50,30)%
}}
\def\realNLOsb{\picb{%
 \Lqu(0,0)(25,15)%
 \Lqu(25,15)(0,30)%
 \Lgl(38,6.6)(50,15)
 \CCirc(25,15){3}{Black}{Black}%
 \TLsc(50,0)(25,15)%
 \TLsc(25,15)(50,30)%
}}
\def\realNLOsc{\picb{%
 \Lqu(0,0)(25,15)%
 \Lqu(25,15)(0,30)%
 \CCirc(25,15){3}{Black}{Black}%
 \TLsc(50,0)(25,15)%
 \TLsc(25,15)(25,25)%
 \TLsc(25,25)(50,40)%
 \Lgl(25,25)(0,40)
}}
\def\realNLOsd{\picb{%
 \Lqu(0,0)(25,15)%
 \Lqu(25,15)(0,30)%
 \CCirc(25,15){3}{Black}{Black}%
 \TLsc(50,-10)(25,5)%
 \TLsc(25,5)(25,15)%
 \TLsc(25,15)(50,30)%
 \Lgl(25,5)(0,-10)
}}
\def\realNLOse{\picb{%
 \Lqu(0,0)(25,15)%
 \Lqu(25,15)(0,30)%
 \CCirc(25,15){3}{Black}{Black}%
 \TLsc(50,0)(25,15)%
 \TLsc(25,15)(25,25)%
 \TLsc(25,25)(0,40)%
 \Lgl(25,25)(50,40)
}}
\def\realNLOsh{\picb{%
 \Lqu(0,0)(25,15)%
 \Lqu(25,15)(0,15)%
 \CCirc(25,15){3}{Black}{Black}%
 \TLsc(0,30)(25,15)%
 \TLsc(25,15)(35,15)%
 \TLsc(35,15)(50,0)%
 \Lgl(35,15)(50,30)
}}
\def\realNLOsi{\picb{%
 \Laqu(50,0)(25,15)%
 \Laqu(25,15)(50,30)%
 \Lgl(12,23.4)(0,15)
 \CCirc(25,15){3}{Black}{Black}%
 \TLsc(0,0)(25,15)%
 \TLsc(25,15)(0,30)%
}}
\def\realNLOsj{\picb{%
 \Laqu(50,0)(25,15)%
 \Laqu(25,15)(50,30)%
 \Lgl(12,6.6)(0,15)
 \CCirc(25,15){3}{Black}{Black}%
 \TLsc(0,0)(25,15)%
 \TLsc(25,15)(0,30)%
}}
\def\realNLOsk{\picb{%
 \Laqu(50,0)(25,15)%
 \Laqu(25,15)(50,30)%
 \CCirc(25,15){3}{Black}{Black}%
 \TLsc(0,0)(25,15)%
 \TLsc(25,15)(25,25)%
 \TLsc(25,25)(0,40)%
 \Lgl(25,25)(50,40)
}}
\def\realNLOsl{\picb{%
 \Laqu(50,0)(25,15)%
 \Laqu(25,15)(50,30)%
 \CCirc(25,15){3}{Black}{Black}%
 \TLsc(0,-10)(25,5)%
 \TLsc(25,5)(25,15)%
 \TLsc(25,15)(0,30)%
 \Lgl(25,5)(50,-10)
}}
\def\realNLOsm{\picb{%
 \Laqu(50,0)(25,15)%
 \Laqu(25,15)(50,30)%
 \CCirc(25,15){3}{Black}{Black}%
 \TLsc(0,0)(25,15)%
 \TLsc(25,15)(25,25)%
 \TLsc(25,25)(50,40)%
 \Lgl(25,25)(0,40)
}}
\def\realNLOsp{\picb{%
 \Laqu(50,0)(25,15)%
 \Laqu(25,15)(50,15)%
 \CCirc(25,15){3}{Black}{Black}%
 \TLsc(50,30)(25,15)%
 \TLsc(25,15)(15,15)%
 \TLsc(15,15)(0,0)%
 \Lgl(15,15)(0,30)
}}
\def\realNLOta{\picb{%
 \Lqu(0,0)(25,15)%
 \Lqu(25,15)(50,0)%
 \CCirc(25,15){3}{Black}{Black}%
 \TLsc(0,15)(12,23.4)%
 \TLsc(12,23.4)(25,15)%
 \TLsc(25,15)(50,30)%
 \Lgl(12,23.4)(0,35) 
}}
\def\realNLOtb{\picb{%
 \Lqu(0,0)(25,15)%
 \Lqu(25,15)(50,0)%
 \CCirc(25,15){3}{Black}{Black}%
 \TLsc(0,30)(25,15)%
 \TLsc(25,15)(38,23.4)%
 \TLsc(38,23.4)(50,15)%
 \Lgl(38,23.4)(50,35)
}}
\def\realNLOtc{\picb{%
 \Lqu(0,0)(25,15)%
 \Lqu(25,15)(50,0)%
 \CCirc(25,15){3}{Black}{Black}%
 \TLsc(0,40)(25,25)%
 \TLsc(25,25)(25,15)%
 \TLsc(25,15)(50,30)%
 \Lgl(25,25)(50,40)
}}
\def\realNLOtd{\picb{%
 \Lqu(0,0)(25,15)%
 \Lqu(25,15)(50,0)%
 \CCirc(25,15){3}{Black}{Black}%
 \TLsc(50,40)(25,25)%
 \TLsc(25,25)(25,15)%
 \TLsc(25,15)(50,30)%
 \Lgl(25,25)(0,40)
}}
\def\realNLOte{\picb{%
 \Lqu(0,0)(25,15)%
 \Lqu(25,15)(50,0)%
 \CCirc(25,15){3}{Black}{Black}%
 \TLsc(50,40)(25,25)%
 \TLsc(25,25)(25,15)%
 \TLsc(25,15)(0,30)%
 \Lgl(25,25)(0,40)
}}
\def\realNLOtf{\picb{%
 \Lqu(0,0)(25,15)%
 \Lqu(25,15)(50,0)%
 \CCirc(25,15){3}{Black}{Black}%
 \TLsc(0,40)(25,25)%
 \TLsc(25,25)(25,15)%
 \TLsc(25,15)(0,30)%
 \Lgl(25,25)(50,40)
}}
\def\realNLOtg{\picb{%
 \Lqu(0,0)(25,15)%
 \Lqu(25,15)(50,0)%
 \CCirc(25,15){3}{Black}{Black}%
 \TLsc(50,40)(25,15)%
 \TLsc(25,15)(40,21)%
 \TLsc(40,21)(50,22)%
 \Lgl(40,21)(0,40)
}}
\def\realNLOth{\picb{%
 \Laqu(50,0)(25,15)%
 \Laqu(25,15)(0,0)%
 \CCirc(25,15){3}{Black}{Black}%
 \TLsc(0,40)(25,15)%
 \TLsc(25,15)(10,21)%
 \TLsc(10,21)(0,22)%
 \Lgl(10,21)(50,40)
}}
\def\extrNLOa{\picb{%
 \Lqu(0,0)(24,14.5)%
 \Lqu(24,15.5)(0,30)%
 \TLsc(24,11)(32,11)%
 \TLsc(24,11)(24,19)%
 \TLsc(32,11)(32,19)%
 \TLsc(24,19)(32,19)%
 \Lgl(32,15)(50,15)%
 \Text(34,15)[c]{$*$}
}}
\def\extrNLOb{\picb{%
 \Laqu(50,0)(26,14.5)%
 \Laqu(26,15.5)(50,30)%
 \TLsc(26,11)(18,11)%
 \TLsc(26,11)(26,19)%
 \TLsc(18,11)(18,19)%
 \TLsc(26,19)(18,19)%
 \Lgl(0,15)(18,15)%
 \Text(0,15)[c]{$*$}
}}
\def\extrNLOc{\picb{%
 \Lqu(0,0)(24,14.5)%
 \Lqu(24,15.5)(0,30)%
 \TLsc(24,11)(32,11)%
 \TLsc(24,11)(24,19)%
 \TLsc(32,11)(32,19)%
 \TLsc(24,19)(32,19)%
 \Lgl(32,15)(50,15)%
 \Text(29,15)[c]{$*$}
 \TLsc(70,0)(50,15)%
 \TLsc(50,15)(70,30)%
}}
\def\extrNLOd{\picb{%
 \Laqu(70,0)(46,14.5)%
 \Laqu(46,15.5)(70,30)%
 \TLsc(46,11)(38,11)%
 \TLsc(46,11)(46,19)%
 \TLsc(38,11)(38,19)%
 \TLsc(46,19)(38,19)%
 \Lgl(38,15)(20,15)%
 \Text(20,15)[c]{$*$}
 \TLsc(0,0)(20,15)%
 \TLsc(20,15)(0,30)%
}}
\def\extrNLOe{\picb{%
 \Lqu(0,0)(21,12)%
 \Lqu(29,12)(50,0)%
 \TLsc(29,11)(21,11)%
 \TLsc(29,11)(29,19)%
 \TLsc(21,11)(21,19)%
 \TLsc(29,19)(21,19)%
 \TLsc(50,50)(25,35)%
 \TLsc(25,35)(0,50)
 \Lgl(25,35)(25,19)%
 \Text(22,20)[c]{$*$}
}}
\def\extrNLOeX{\picb{%
 \Lqu(0,0)(25,15)%
 \Lqu(25,15)(50,0)%
 \CCirc(25,15){3}{Black}{Black}%
 \TAsc(25,25)(9,0,360)
 \TLsc(50,60)(25,45)%
 \TLsc(25,45)(0,60)
 \Lgl(25,45)(25,35)%
 \Text(17.5,36)[c]{\tiny$\scriptstyle\rm V$}
 \Text(16,26.5)[r]{\tiny$\scriptstyle\rm V$}
 \Text(26,17)[l]{\tiny$\,\sim\! T^2_{ }$}
 \Text(21,28)[c]{\tiny$*$}
}}
\def\vertNLOa{\picb{%
 \Lqu(0,0)(24,14.5)%
 \Lqu(24,15.5)(0,30)%
 \Lgl(27.5,15)(47.5,15)%
 \CBoxc(25,15)(5,5){Black}{Black}%
}}
\def\vertNLOb{\picb{%
 \Lqu(0,-10)(16,2)%
 \Lqu(16,28)(0,40)%
 \TLsc(16,2)(16,28)%
 \Lgl(16,28)(33,15)%
 \Lgl(16,2)(33,15)%
 \Lgl(33,15)(53,15)%
 \Text(6,10.2)[c]{\tiny $\ell^{ }_a$}%
 \Text(19,20)[c]{\tiny $W$}%
 \Text(19,0)[c]{\tiny $W$}%
 \Line(-10,-15)(-5,-15)%
 \Line(-10,-15)(-10,45)%
 \Line(-10,45)(-5,45)%
}}
\def\vertNLOc{\picb{%
 \Lqu(0,-10)(16,2)%
 \Lqu(16,28)(0,40)%
 \Lgl(16,2)(16,28)%
 \TLsc(16,28)(33,15)%
 \TLsc(16,2)(33,15)%
 \Lgl(33,15)(53,15)%
 \Text(6,10.2)[c]{\tiny $W$}%
 \Text(19,20)[c]{\tiny $\ell^{ }_a$}%
 \Text(19,0)[c]{\tiny $\ell^{ }_a$}%
}}
\def\vertNLOd{\picb{%
 \Lqu(0,0)(25,15)%
 \Lqu(25,15)(0,30)%
 \Lgl(25,15)(40,15)%
 \Asc(50,15)(10,0,360)%
 \Lgl(60,15)(75,15)%
 \Text(22,5)[c]{\tiny $Z$}%
 \Text(48,5)[c]{\tiny $\gamma$}%
 \Text(28,29)[l]{\tiny any charged}%
 \Text(28,23)[l]{\tiny particle}%
}}
\def\vertNLOe{\picb{%
 \Lqu(0,0)(25,15)%
 \Lqu(25,15)(0,30)%
 \Lgl(25,15)(40,15)%
 \Line(36,11)(44,19)%
 \Line(36,19)(44,11)%
 \Lgl(40,15)(55,15)%
 \Line(60,-15)(55,-15)%
 \Line(60,-15)(60,45)%
 \Line(60,45)(55,45)%
 \Text(45,-9)[l]{\tiny full,IR}%
}}
\def\vertNLOf{\picb{%
 \Lqu(0,0)(25,15)%
 \Lqu(25,15)(0,30)%
 \CCirc(25,15){3}{Black}{Black}%
 \TAsc(35,15)(10,0,360)%
 \Lgl(45,15)(60,15)%
 \Line(-10,-15)(-5,-15)%
 \Line(-10,-15)(-10,45)%
 \Line(-10,45)(-5,45)%
 \Line(70,-15)(65,-15)%
 \Line(70,-15)(70,45)%
 \Line(70,45)(65,45)%
 \Text(51,-9)[l]{\tiny eft,IR}%
}}
\makeatletter \@addtoreset{equation}{section} \makeatother
\renewcommand{\theequation}{\arabic{section}.\arabic{equation}}
\makeatletter
\renewcommand\section{\@startsection {section}{1}{\z@}%
                                   {-5.5ex \@plus -1ex \@minus -.2ex}
                                   {2.3ex \@plus.2ex}%
                                   {\normalfont\large\bfseries}}
\renewcommand\subsection{\@startsection{subsection}{2}{\z@}%
                                     {-3.25ex\@plus -1ex \@minus -.2ex}%
                                     {1.5ex \@plus .2ex}%
                                     {\normalfont\normalsize\bfseries}}
\renewcommand\thesection {\@arabic\c@section}
\renewcommand\thesubsection   {\thesection.\@arabic\c@subsection}
\renewcommand{\@seccntformat}[1]{%
\csname the#1\endcsname.\hspace{1.0em}}
\makeatother

\begin{document}

\flushbottom

\begin{titlepage}

\begin{flushright}
March 2024
\end{flushright}
\begin{centering}
\vfill

{\Large{\bf
     QED corrections to the thermal neutrino interaction rate
}} 

\vspace{0.8cm}

G.~Jackson$^\rmi{a,b}$ and 
M.~Laine$^\rmi{c}$,

\vspace{0.6cm}

${}^\rmi{a}_{ }${\em
Institute for Nuclear Theory,
Box 351550,  
University of Washington, \\ 
Seattle, WA 98195-1550, United States \\}

\vspace*{0.3cm}

${}^\rmi{b}_{ }${\em
SUBATECH,
Nantes Universit\'e, 
IMT Atlantique, 
IN2P3/CNRS, \\ 
4 rue Alfred Kastler, 
La Chantrerie BP 20722, 
44307 Nantes, France \\}

\vspace*{0.3cm}

${}^\rmi{c}_{ }${\em
AEC, 
Institute for Theoretical Physics, 
University of Bern, \\ 
Sidlerstrasse 5, CH-3012 Bern, Switzerland \\}

\vspace*{0.6cm}

{\em 
Emails: jackson@subatech.in2p3.fr, laine@itp.unibe.ch}

\vspace*{0.8cm}

\mbox{\bf Abstract}
 
\end{centering}

\vspace*{0.3cm}
 
\noindent
Motivated by precision computations of neutrino decoupling at MeV 
temperatures, we show how QED corrections to the thermal neutrino interaction
rate can be related to the electron-positron spectral function as well as 
an effective $\bar{\nu}\nu\gamma$ vertex. The spectral function is needed 
both in a timelike and in a spacelike domain, and for both of its physical 
polarization states (transverse and longitudinal with respect to spatial 
momentum). Incorporating an NLO evaluation of this spectral function, 
an estimate of the $\bar{\nu}\nu\gamma$ vertex, and HTL resummation 
of scatterings mediated by soft Bose-enhanced $t$-channel photons,  
we compute the interaction rate as a function of the neutrino momentum 
and flavour. Effects on the $ -(0...2)\%$ level are found, noticeably 
smaller than a previous estimate of a related quantity.

\vfill


\end{titlepage}

\tableofcontents

%
\section{Introduction}
\la{se:intro}

The energy density carried by relativistic 
degrees of freedom at the time of primordial nucleosynthesis 
or photon decoupling is parametrized by an effective number of neutrino 
species, $N^{ }_\rmi{eff}$. It can be inferred experimentally, and 
computed theoretically, either in the Standard Model, or in a given 
extension thereof. In the Standard Model, the naive estimate is~3, 
but the precisely computed 
value differs from this. The current best estimate, 
obtained after many decades of work, reads 
$
 N^\rmii{(SM)}_\rmi{eff} \approx 3.043 \pm 0.001
$~\cite{Neffm2,Neffm1,Neff0,cemp}. 
Were the observed value to eventually differ from $N^\rmii{(SM)}_\rmi{eff}$, 
this could be an indication of physics beyond the Standard Model. 

The latest ingredient included in the Standard Model prediction
of $N^{ }_\rmi{eff}$ originates
from QED corrections to the rate at which neutrinos interact with the 
electron-positron plasma at the time of their decoupling, 
$T\sim (1 ... 3)$\hspace*{0.3mm}MeV~\cite{cemp}. This result was
based on previous work in a different 
(astrophysical) environment~\cite{old}. 
Concretely, the observable evaluated is 
\be
 Q \; \equiv \; \frac{\delta \rho}{\delta t} 
 \biggr|^{ }_{e^+_{ } e^-_{ } \to \bar{\nu}\nu(\gamma)}
 \; \equiv \; 
 \bigl\langle  
 \,( \epsilon^{ }_{e^+_{ }} + \epsilon^{ }_{e^-_{ }} )\,
 {\textstyle \sum} 
 |\mathcal{M}|^2_{ e^+_{ } e^-_{ } \to \bar{\nu} \nu(\gamma) }
 \bigr\rangle^{ }_\rmi{phase~space}
 \;, \la{def_Q}
\ee
where $\rho$ denotes the energy density. The last form 
stands for the phase-space average of 
the electron-positron energy loss,
weighted by the matrix element squared of this process. 

Given that the computation of refs.~\cite{cemp,old} is quite complicated, 
and that it entails certain approximations, it seems 
well motivated to carry out an independent analysis of 
the rate at which neutrinos interact with the QED plasma. 
However, there is also 
the conceptual issue that observables should in principle
be defined without reference to Boltzmann equations, 
which have a limited range of applicability. 
Therefore, we consider a quantity
which is related to \eq\nr{def_Q} but {\em not}
equivalent, with the benefit that  
it can be unambiguously defined beyond leading order.
Concretely, this is achieved
by considering the imaginary part 
of the retarded neutrino self-energy, 
frequently called the thermal neutrino interaction rate. 
Physically, this quantity can be argued to affect 
the time evolution of a neutrino density matrix
(cf.\ \se\ref{ss:observables}). 

Our computation entails some practical differences
compared with refs.~\cite{cemp,old}. First of all, 
$e^+_{ }e^-_{ }\to\bar{\nu}\nu(\gamma)$ represents only a subclass
of the processes that we consider (cf.\ \fig\ref{fig:processes}
on page~\pageref{fig:processes} for the full set);
we add reactions like  
$e\nu \to e\nu$ including its 1-loop virtual corrections, 
or $e\gamma\to e\bar{\nu}\nu$, 
or a logarithm $\sim\ln(T/m^{ }_e)$
that can be associated with a loop-induced 
$\bar{\nu}\nu\gamma$ vertex. 
Second, we include Pauli blocking of the final-state neutrinos, 
and Bose enhancement of both real and virtual photons, 
and Hard Thermal Loop (HTL) resummation required for
properly treating soft virtual photons, which leads to 
another logarithm $\sim \ln(1/\alpha^{ }_\rmi{em})$. 
Third, \eq\nr{def_Q} represents a momentum average, whereas we 
determine the differential interaction rate, as a function of 
momentum ($k$) with respect to the heat bath. 
But there are simplifications on our side as well, 
notably we assume that the neutrino and electron ensembles 
carry the same temperature, which is physically the case
only at the start of the decoupling process; and we omit
the electron mass, which is again best justified at the 
highest temperatures.  

In view of these differences, both in the observable itself
and on the technical side, it may not be surprising that our
results differ from those 
in ref.~\cite{cemp} (cf.\ \se\ref{se:concl}). 
Notably, restricting to the so-called $s$-channel processes, 
corresponding to those evaluated in ref.~\cite{cemp}, 
we find the opposite sign of the relative NLO correction;
a significantly smaller overall magnitude;
and a larger dependence on the neutrino flavour. 

Our presentation is organized as follows. 
In \se\ref{se:setup}, we describe our starting point, 
the essential methods used, 
and the main steps of our computation.
A reader not interested in such details is 
advised to skip directly to \se\ref{se:nlo},  
where our NLO results are given and elaborated upon. 
Conclusions are offered in \se\ref{se:concl}. Appendix~A 
displays computational details related to the leading-order (LO) evaluation, 
appendix~B an HTL-resummed computation of the leading-logarithmic
part of the NLO result, and appendix~C supplementary plots. 

%
\section{Technical setup}
\la{se:setup}

The computation of this paper is technical in nature, and in this
section we specify the steps that it entails. Our tool is thermal
field theory in the imaginary-time formalism. Parts of the results may
also be obtained from Boltzmann equations, however the latter are 
not well suited to incorporating temperature-dependent virtual effects, 
even though at the NLO level virtual effects can be as important as 
real corrections (and cancel parts of them). 

%
\subsection{Observables}
\la{ss:observables}

To describe the non-equilibrium time evolution of neutrinos 
in the early universe, it is useful to define 
a density matrix for a momentum mode~$k \equiv |\vec{k}|$,
\be
 \rho^{ }_{ab} \; \equiv \; e^{i \phi^{ }_{ab}(t)}_{ } 
 w^\dagger_a w^{ }_b  
 \;, \la{hat_rho_1}
\ee
where $w^\dagger_{a}$ and $w^{ }_b$ are creation and annihilation operators, 
respectively, and $a,b$ are flavour indices.\footnote{%
 Often the flavour indices are placed in the opposite order, to 
 have a text-book appearance of the overall 
 sign in the first term of \eq\nr{vn_1}~\cite{sr}.
 } 
The phase factor 
$\phi^{ }_{ab}$ depends on the time evolution picture chosen. 
After averaging over the ``fast'' medium degrees of freedom, 
the effective evolution equation for the ``slow'' variables
is sometimes assumed to take the form
\be
 \dot{\rho} \;\simeq\;
 i \bigl[\,\Delta\mathcal{E}^{ }_k\,,\,\rho \,\bigr] 
 + \frac{1}{2} \bigl\{ \Gamma^{ }_k\,,\, \rho^{ }_\rmi{eq} - \rho \bigr\}
 \;. \la{vn_1}
\ee
Here the time derivative reads 
$
 \dot{(...)} \equiv (\partial^{ }_t - H k \partial^{ }_k)(...)
$, 
where $H$ is the Hubble rate; 
$\Delta\mathcal{E}^{ }_k$ is a matrix involving energy differences;  
and $ \rho^{ }_\rmi{eq} $ is a would-be equilibrium 
density matrix. 
Equation~\nr{vn_1} can be obtained from linear response theory, 
if the density matrix is perturbed around equilibrium 
in a single momentum bin~$k$. 
In this language, our goal 
is to determine the matrix $\Gamma^{ }_k$.\footnote{%
  Another important effect are medium modifications
  to $\Delta\mathcal{E}^{ }_k$~\cite{raffelt}. 
  They are of $\rmO(G^{ }_\rmiii{F})$, where 
  $G^{ }_\rmiii{F}$ denotes the Fermi coupling, 
  whereas $\Gamma^{ }_k$ is of $\rmO(G^{2}_\rmiii{F})$.
 } 
We denote its diagonal components 
(in the interaction basis)
by 
$
 \Gamma^\rmii{ }_{k;\nu^{ }_a}
$.

Actually, the form sketched in \eq\nr{vn_1} is oversimplified.
Through weak interactions, the neutrinos 
interact with electrons and positrons 
through pair production, 
pair annihilation, and elastic scattering.  
The temperature of the 
electron-positron plasma 
is denoted by~$T$. 
However, the neutrinos also experience 
interactions with each other.
As they are falling out of equilibrium, the effective temperatures of 
the neutrino ensembles start to differ from~$T$. In this situation, 
the scatterings off neutrinos 
drive the neutrinos not towards a universal equilibrium
distribution, but rather towards 
the scatterers' ensemble.
Having this in mind, we disentagle the full interaction rate 
into partial ones, 
\be
 \Gamma^{ }_{k;\nu^{ }_a} 
 \;\simeq\;
 \sum_i
 \Gamma^{(i)}_{k;\nu^{ }_a} 
 \;, \la{vn_2}
\ee
where $i$ labels the different ensembles. 
We employ this representation in \se\ref{se:nlo}, however 
assume the presence of a universal temperature in all the ensembles. 

%
\subsection{Notation for the practical computation}
\la{ss:notation}

In order to avoid excessive imaginary units and minus signs, 
we employ Euclidean conventions for the practical computation, 
with the path integral weight appearing as 
\be
 e^{\;i \mathcal{S}^{ }_\M}_{ } \longrightarrow e^{-S^{ }_\E}_{}
 \;.
\ee
Minkowskian ($t$) and Euclidean ($\tau$) 
time coordinates are formally related via the Wick rotation
$\tau \leftrightarrow i t$, whereas for the corresponding momenta
we have 
$ k^{ }_n \leftrightarrow -i k^0_{ }$, 
where the Matsubara frequencies read $k^{ }_n = 2 n \pi T $ for
bosons, and $k^{ }_n = (2n + 1)\pi T$ for fermions, with $n \in \mathbbm{Z}$.
Whenever in Minkowskian spacetime, the signature ($+$$-$$-$$-$) is 
employed. Minkowskian four-momenta are denoted by 
$\K = (k^0_{ },\vec{k})$, 
Euclidean ones by 
$K = (k^{ }_n,\vec{k})$. 
We use an implicit notation where the symbol used 
indicates whether analytic continuation has been carried out, 
for example
\be
 \Sigma^{ }_{\K} 
 \; \equiv \; 
 \Sigma^{ }_{K} \bigr|^{ }_{k^{ }_n\to -i[k^0_{ } + i 0^+_{ }]}
 \;, \quad
 \Sigma^{ }_k 
 \; \equiv \; 
 \Sigma^{ }_{\K} \bigr|^{ }_{k^0_{ }\to k}
 \;. 
\ee

Euclidean Dirac matrices, identified by the fact 
that all indices are down, are defined as 
\be
 \gamma^{ }_0 \; \equiv \; \gamma^0_{ }
 \;, \quad
 \gamma^{ }_i \; \equiv \; - i \gamma^i_{ }
 \;. 
\ee
It follows that 
\be
 A \gamma^{ }_\mu B \, C \gamma^{ }_\mu D = 
 A \gamma^{ }_\mu B \, C \gamma^\mu_{ } D
 \;, \quad
 i \bsl{K} = \bsl{\K} 
 \;. \la{rules}
\ee
The Fermi coupling reads at tree level, or leading order (LO),
\be
 G^{ }_\rmii{F} \;\stackrel{\rmii{LO}}{\equiv}\; \frac{1}{\sqrt{2} v^2}
                \;\stackrel{\rmii{LO}}{\equiv}\; \frac{g_2^2}{4\sqrt{2} \mW^2}
 \;, \quad v \simeq 246~\mbox{GeV}
 \;, 
\ee
and the Weinberg angle is parametrized by 
\be
 \xW \equiv \sin^2 \theta^{ }_\aW 
            \;\stackrel{\rmii{LO}}{\equiv}\; \frac{g_1^2}{g_1^2 + g_2^2}
 \;, \la{xW}
\ee
where $g^{ }_1$ and $g^{ }_2$ are the gauge couplings of 
U$^{ }_\rmii{Y}$(1) and SU$^{ }_\rmii{L}$(2), respectively.
As we are carrying out a loop computation, the precise definitions
of $G^{ }_\rmii{F}$ and $\xW$
need to be revisited later on. 
The numerical values used for their $\msbar$
renormalized versions at the scale $\bmu = m^{ }_e$ are specified 
in the next section. 

%
\subsection{Interactions according to a Fermi effective theory}
\la{ss:fermi}

At temperatures $T \sim $~MeV, all hadrons, as well as the $\mu$ and 
$\tau$ leptons, are heavy, appearing in the thermal bath with 
an exponentially suppressed weight. Therefore we only need to include
neutrinos, electrons, positrons and photons 
as dynamical degrees of freedom. Working in 
the interaction basis, we may thus take the Lagrangian
\ba
 L^{ }_\E & \supset & \frac{G^{ }_\rmii{F}}{4\sqrt{2}}
 \biggl\{ \,
 \bar{\nu}^{ }_a \gamma^{ }_\mu (1-\gamma^{ }_5) \nu^{ }_a 
 \, \bar{\nu}^{ }_b  \gamma^{ }_\mu (1 - \gamma^{ }_5) \nu^{ }_b
 \; + \;  
    4\, \bar{\nu}^{ }_e \gamma^{ }_\mu (1-\gamma^{ }_5) \ell^{ }_e 
    \, \bar{\ell}^{ }_e  \gamma^{ }_\mu (1 - \gamma^{ }_5) \nu^{ }_e 
 \nn & - & 
  2\,  
 \bar{\nu}^{ }_a \gamma^{ }_\mu (1-\gamma^{ }_5) \nu^{ }_a 
    \, \bar{\ell}^{ }_e  \gamma^{ }_\mu 
           \bigl(1 - 4 \xW - \gamma^{ }_5\bigr) \ell^{ }_e 
 \biggr\}
 \; + \; 
 \delta L^{ }_\E
 \;,  
 \la{S_E_1}
 \\
 \delta L^{ }_\E & \supset & 
 \frac{ i e G^{ }_\rmii{F} }{4\sqrt{2}}  
 \, C^{ }_a \, 
 \bar{\nu}^{ }_a \gamma^{ }_\mu (1-\gamma^{ }_5) \nu^{ }_a 
 \,\partial^{ }_\nu F^{ }_{\nu\mu}
 \;, \la{ct_1}
\ea
as a starting point. Here $a,b$ are flavour indices, and a sum over
repeated indices is implied. The operators in \eq\nr{S_E_1} are 
generated at tree-level, and we have also adopted tree-level values
for the couplings (see below). 
The operator in \eq\nr{ct_1}, where $F^{ }_{\nu\mu}$ 
is the QED field strength and $e$ is the electromagnetic coupling, 
is generated at 1-loop level, 
through the diagrams shown in \fig\ref{fig:vertex}
(cf.,\ e.g., refs.~\cite{eft1,eft2}). 
The coefficient $C^{ }_a$ 
includes a computable divergence and anomalous dimension, 
but its absolute value cannot be determined perturbatively, as it gets
a contribution from hadronic effects through the $Z\gamma$ bubble. 
We return to a discussion of its effects
at the end of this section (cf.\ \eq\nr{C_a}).

%
\begin{figure}[t]
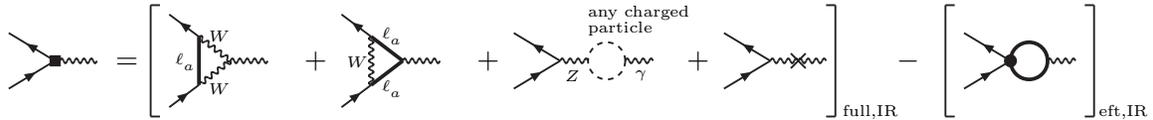


\begin{eqnarray}
&&
 \hspace*{-15mm}
 \vertNLOa \hspace*{1mm} = \hspace*{2mm}  
 \vertNLOb \hspace*{5mm} + 
 \vertNLOc \hspace*{5mm} + 
 \vertNLOd \hspace*{10mm} + 
 \vertNLOe \hspace*{10mm} - \hspace*{4mm}
 \vertNLOf
 \nonumber
\end{eqnarray}

\vspace*{3mm}

\caption[a]{\small 
 Standard Model processes generating
 the effective $\bar\nu^{ }_a \nu^{ }_a \gamma$
 vertex in \eq\nr{ct_1}. 
 A cross denotes a counterterm,  
 IR indicates that the amplitude is evaluated 
 in a kinematic domain in which the effective theory (eft) is applicable, 
 and a thick solid line in the eft contribution
 stands for an electron or positron. 
 The subtraction removes low-energy contributions, 
 which are regenerated 
 in a more complete form
 through \fig\ref{fig:processes}
 (including thermal corrections).
} 
\la{fig:vertex}
\end{figure}
%

The representation in \eq\nr{S_E_1} can be brought into a more
helpful form by making use of a Fierz identity. This is particularly 
transparent, if we work in the Weyl representation of the Euclidean
Dirac matrices, 
\be 
 \gamma^{ }_0 = 
 \left( 
   \begin{array}{cc}
         0 & \mathbbm{1} \\ 
         \mathbbm{1} & 0 
   \end{array} 
 \right)
 \;, \quad 
 \gamma^{ }_m = 
 \left( 
   \begin{array}{cc}
         0 & -i\sigma^{ }_m \\ 
         i \sigma^{ }_m & 0 
   \end{array} 
 \right)
 \;, \quad 
 \mathbbm{1} - \gamma^{ }_5 = 
  \left( 
   \begin{array}{cc}
         2\times\mathbbm{1} & 0 \\
         0 & 0  
   \end{array} 
 \right)
 \;.
\ee
Inserting 
$ 
 \sum_m^{ }
 (\sigma^{ }_m)^{ }_{ij} (\sigma^{ }_m)^{ }_{kl} 
 = 2 \delta^{ }_{il}\delta^{ }_{jk} - \delta^{ }_{ij}\delta^{ }_{kl}
$, 
this gives
\be
 [\gamma^{ }_\mu\,(1-\gamma^{ }_5)]^{ }_{ij}
 [\gamma^{ }_\mu\,(1-\gamma^{ }_5)]^{ }_{kl}
 \; = \; 
 8
  \left( 
   \begin{array}{cc}
         0 & 0 \\
         \delta^{ }_{ij}\delta^{ }_{kl}
  - \delta^{ }_{il} \delta^{ }_{jk} & 0  
   \end{array} 
 \right)
 \;. 
\ee
This structure is antisymmetric in $j \leftrightarrow l$. This is 
compensated for by another minus sign from the anticommutation of 
fermionic fields, whereby \eq\nr{S_E_1} can finally be turned into
\ba
 L^{ }_\E & \supset & \frac{G^{ }_\rmii{F}}{4\sqrt{2}}
 \biggl\{ \,
 \bar{\nu}^{ }_a \gamma^{ }_\mu (1-\gamma^{ }_5) \nu^{ }_a 
 \, \bar{\nu}^{ }_b  \gamma^{ }_\mu (1 - \gamma^{ }_5) \nu^{ }_b
 \nn & + & 
  2\,  
 \bar{\nu}^{ }_{a} \gamma^{ }_\mu (1-\gamma^{ }_5) \nu^{ }_{a} 
    \, \bar{\ell}^{ }_e  \gamma^{ }_\mu 
           \bigl[ 
                   2\delta^{ }_{a,e} - 1 + 4 \xW
                 + (1 - 2\delta^{ }_{a,e} )\gamma^{ }_5
           \bigr] \ell^{ }_e 
 \biggr\}
 \; + \; 
 \delta L^{ }_\E
 \;. 
 \la{S_E_2}
\ea
This is the form that we employ for our practical computations. 

Let us now return to the operator 
$\delta L^{ }_\E$, defined in \eq\nr{ct_1}. 
It can be written in the same form as the operators in \eq\nr{S_E_2} 
by making use of equations of motion~(eom), leading to 
\be
 \delta L^{ }_\E \bigr|^{ }_\rmi{eom} \; \supset \;
 \frac{G^{ }_\rmii{F}}{4\sqrt{2}}
 \biggl\{ \,
  e^2_{ } C^{ }_a \,
 \bar{\nu}^{ }_{a} \gamma^{ }_\mu (1-\gamma^{ }_5) \nu^{ }_{a} 
    \, \bar{\ell}^{ }_e  \gamma^{ }_\mu  \ell^{ }_e 
 \biggr\}
 \;. \la{ct_2}
\ee
Thereby \eq\nr{ct_2} can be viewed as an $\rmO(e^2)$ correction
to the operators in \eq\nr{S_E_2}. Indeed, 
ref.~\cite{eft2} determined the NLO
coefficients of the 4-fermion operators after this redefinition,
by fixing the low-energy hadronic contribution to the $Z\gamma$ bubble from 
experimentally measured quantities~\cite{eft3}. However, the goal 
of the present paper is to compute the percentual change 
from QED corrections. Therefore, it is more transparent to keep 
the coefficients of the 4-fermion operators at their electroweak
values, and track the QED corrections, such as those originating
from \eq\nr{ct_1}, separately. To achieve this, 
we need to reconstruct the value of~$C^{ }_a$ from the 4-fermion
coefficients estimated in ref.~\cite{eft2}.

In order to achieve this, let us re-express the neutrino-electron interaction
from \eqs\nr{S_E_2} and \nr{ct_2} in the same form as in ref.~\cite{eft2}, 
\ba
 L^{ }_\E + \delta L^{ }_\E & \supset &
 2\sqrt{2} G^{ }_\rmii{F}
 \,
 \bar{\nu}^{ }_{a} \gamma^{ }_\mu \aL \nu^{ }_{a}
 \,
 \bar{\ell}^{ }_e  \gamma^{ }_\mu 
 \, \biggl\{ 
 \nn[3mm] 
 & + &  
    \biggl[ 
     \delta^{ }_{a,e}\, 
     \biggl( 
     \underbrace{
     \xW + \frac{ e^2_{ } C^{ }_a }{8} 
     }_{\,\equiv\,g^{ }_\rmii{$e$R}}
     \biggr)
    + 
     \bigl( 1 - \delta^{ }_{a,e} \bigr)
     \biggl(
     \underbrace{
     \xW + \frac{ e^2_{ } C^{ }_a }{8} 
     }_{\,\equiv\,g^{ }_\rmii{$(\mu,\tau)$R}}
     \biggr)
     \biggr]
    \, \aR
 \nn[3mm]
 & + &  
    \biggl[ 
      \delta^{ }_{a,e}\, 
        \biggl( 
        \underbrace{
        \xW + \fr12 + \frac{ e^2_{ } C^{ }_a }{8} 
        }_{\,\equiv\,g^{ }_\rmii{$e$L}}
        \biggr)
    + 
     \bigl( 1 - \delta^{ }_{a,e} \bigr)
     \biggl(
        \underbrace{
        \xW - \fr12 + \frac{ e^2_{ } C^{ }_a }{8}
        }_{\,\equiv\,g^{ }_\rmii{$(\mu,\tau)$L}}
     \biggr)
    \biggr]\, \aL
 \, \biggr\} 
 \, \ell^{ }_e 
 \;, \la{ct_3}  
 \hspace*{8mm}
\ea
where $\aR \equiv (1+\gamma^{ }_5)/2$
and $\aL \equiv (1-\gamma^{ }_5)/2$ are 
the right and left-handed projectors. 
The coefficients 
$g^{ }_\rmii{$e$R},...,g^{ }_\rmii{$(\mu,\tau)$L}$ 
correspond to a notation employed 
in the literature
(cf.,\ e.g.,\ ref.~\cite{mea}), 
where often no distinction is made between $a=\mu$ and $a=\tau$.

After renormalization, the finite parts of the 
coefficients $g^{ }_\rmii{$a$R}$ 
and $g^{ }_\rmii{$a$L}$ become running parameters. 
We fix them through their values 
at the $\msbar$ scheme renormalization scale $\bmu = m^{ }_e$.
The logarithmic running, induced by $C^{ }_a$ 
(cf.\ \eq\nr{C_a}), 
matches a corresponding thermal logarithm, and ultimately 
yields the $\ln(m^{ }_e/T)$ visible in \eq\nr{ABCD}.

To be concrete, the coefficients 
$
 2\sqrt{2} G^{ }_\rmii{F}\, g^{ }_\rmii{$a$R,L} 
$
from \eq\nr{ct_3} 
agree with the coefficients 
$
 c^{\nu_{\!\ell}^{ } \ell'}_\rmii{R,L} 
$
listed in table~4 of ref.~\cite{eft2}.
Relevant for us is  
the case that only the charged flavour $\ell' = \ell^{ }_e$ is light.
In fact, the comparison is overconstrained, as we have parametrized 
six independent coefficients 
$
 c^{\nu_a{ } \ell_e^{ }}_\rmii{R,L} 
$, 
$a\in\{e,\mu,\tau\}$,
through the three parameters~$C^{ }_a$, 
and thereby omitted some (non-QED) electroweak corrections. 
Nevertheless, all values
from ref.~\cite{eft2} can be conservatively encompassed by 
setting 
$
 G^{ }_\rmii{F} |^{ }_{\bmu = m^{ }_e} \simeq 1.1664 \times 10^{-5}_{ }  
 / \mbox{GeV}^2_{ }
$, 
$
 \xW |^{ }_{\bmu = m^{ }_e} \simeq 0.2386
$, 
and 
\be
 \frac{ e^2_{ }C^{ }_a }{8} \biggr|^{ }_\rmi{bare} 
 \; = \;
 ( 2\delta^{ }_{a,e} - 1 + 4 \xW )
 \frac{e^2}{3} \frac{\mu^{-2\epsilon}}{(4\pi)^2} 
 \biggl( \frac{1}{\epsilon} + \ln\frac{\bmu^2}{m_e^2} \biggr)
 \; + \;
 (-0.01 \, ... \, 0.01)
 \;. \la{C_a}
\ee
The first term contains the 
correct divergence (i.e.\ counterterm)
and renormalization scale dependence; 
the last parentheses represent 
the uncertainty from our simplified procedure. 
We remark that this uncertainty is of the order of the smallest
physical coupling entering our computation, cf.\ \eq\nr{gmuV}.

An important point to appreciate is that 
QED physics is more transparent in the original
vector--axial rather than in the new left--right basis.
After taking care of renormalization 
in the left--right basis, we can transfer back
to the vector--axial basis,
through
\be
       g^{ }_\rmii{$a$R} \aR + 
       g^{ }_\rmii{$a$L} \aL
       \;=\; 
       \underbrace{
       \frac{ g^{ }_\rmii{$a$R} + g^{ }_\rmii{$a$L}}{2}
       }_{\equiv\; g^{ }_\rmii{$a$V}}
       \;+\; 
       \underbrace{
       \frac{ g^{ }_\rmii{$a$R} - g^{ }_\rmii{$a$L}}{2}
       }_{\equiv\; g^{ }_\rmii{$a$A}}
       \,\gamma^{ }_5
 \;. 
\ee 
One reason for the simplification
is that QED is a vectorlike theory, 
and therefore the QED running, induced by $C^{ }_a$, 
affects the vector coupling, 
but cancels in the axial one, cf.\ \eq\nr{ct_2}.
Another reason is that the vector coupling is very 
small for $a\neq e$,
\be
 g^{ }_\rmii{$(\mu,\tau)$V} \; = \; 
 \xW - \frac{1}{4} + 
\frac{ e^2_{ } C^{ }_{(\mu,\tau)} }{8} 
 \;  
   \overset{ \rmii{ $\msbar$ } }{  
  \underset{ \rmii{$\;\;\bmu = m^{ }_e$} }{ \simeq } } 
 \; 
 -0.01 
 \; + \;
 (-0.01 \, ... \, 0.01)
 \;. \la{gmuV}
\ee
A virtual electron-positron loop couples to a photon
only through $g^{ }_\rmii{$a$V}$. Therefore
the flavours $a\neq e$ are almost insensitive 
to scatterings mediated by photon exchange.
As we will see,   
this induces a significant difference to the
thermal interaction rates felt
by $\nu^{ }_e$ and $\nu^{ }_{a\neq e}$.

%
\subsection{Contractions for neutrino self-energy}
\la{ss:wick}

The next task is to determine 
the medium-modified (i.e.\ full) neutrino propagator. 
After resumming 1-loop self-energy
insertions, and showing explicitly the chiral projectors
in accordance with the appearance of only left-handed
neutrinos in \eq\nr{S_E_2}, 
the inverse propagator is expressed as 
\be
 \Delta^\rmi{(full)}_{K;\nu^{ }_a}
 \; = \;
 \aR \, \bigl( \,  
 i \bsl{K} + \Sigma^{ }_{K;\nu^{ }_a}
 \, \bigr) \, \aL
 \;. \la{prop_full}
\ee
If we furthermore define the electron currents
\be
 J^{ }_\mu \; \equiv \; 
   \bar{\ell}^{ }_e \gamma^{ }_\mu \ell^{ }_e
 \;, \quad
 J^{5}_\mu \; \equiv \; 
  \bar{\ell}^{ }_e \gamma^{ }_\mu \gamma^{ }_5\, \ell^{ }_e
 \;, 
\ee
and their fermion-line connected (``c'') correlation functions 
\ba
 V^{ }_{P;\mu\nu} 
 & \equiv &
 \int_X e^{i P\cdot X}_{ }
 \bigl\langle\, J^{ }_\mu(X) J^{ }_\nu(0) \,\bigr\rangle^{ }_\rmi{c} 
 \;, \la{Vmunu} \\ 
 A^{ }_{P;\mu\nu} 
 & \equiv &
 \int_X e^{i P\cdot X}_{ }
 \bigl\langle\, J^{5}_\mu(X) J^{5}_\nu(0) \,\bigr\rangle^{ }_\rmi{c} 
 \;, \la{Amunu} \\ 
 M^{ }_{P;\mu\nu} 
 & \equiv &
 \int_X e^{i P\cdot X}_{ }
 \bigl\langle\, J^{ }_\mu(X) J^{5}_\nu(0)
 + J^{5}_\mu(X) J^{ }_\nu(0) \,\bigr\rangle^{ }_\rmi{c} 
 \;, \la{Mmunu}
\ea
then the $\rmO(G_\rmii{F}^2)$ contribution to the self-energy becomes 
\ba
 \Sigma^{ }_{K;\nu^{ }_a} 
 & \supset & 
 \frac{G_\rmii{F}^2}{8} 
 \gamma^{ }_\mu (1-\gamma^{ }_5) 
 \Delta^{-1}_{K+P;\nu^{ }_a}
 \mathbbm{S}^{a}_{P;\mu\nu}
 \,  \gamma^{ }_\nu (1-\gamma^{ }_5)  
 \;, \la{Sigma_full}
 \\[3mm]
 \mathbbm{S}^{a}_{P;\mu\nu}
 & = & 
 {\textstyle \sum_b}
 \tr \bigl[\, 
   \Delta^{-1}_{P+Q;\nu^{ }_b} \gamma^{ }_\mu ( 1-\gamma^{ }_5 )
   \Delta^{-1}_{Q;\nu^{ }_b} \gamma^{ }_\nu (1-\gamma^{ }_5)
 \,\bigr] 
 \nn[2mm] 
 & - & 
 \gamma^{ }_\nu (1-\gamma^{ }_5)
   \Delta^{-1}_{P+Q;\nu^{ }_a} \gamma^{ }_\mu ( 1-\gamma^{ }_5 )
   \Delta^{-1}_{Q;\nu^{ }_a} 
 \nn[2mm] 
 & - & 
 \, \Bigl[ \,
   \bigl( 2\delta^{ }_{a,e} - 1 + 4 \xW \bigr)^2_{ }\, V^{ }_{P;\mu\nu}
 + 
   \bigl( 1 - 2\delta^{ }_{a,e} \bigr)
   \bigl( 2\delta^{ }_{a,e} - 1 + 4 \xW \bigr) M^{ }_{P;\mu\nu}
 + A^{ }_{P;\mu\nu}
 \, \Bigr] 
 \nn[2mm]
 & + & 
 \, e^2 \bigl( 2\delta^{ }_{a,e} - 1 + 4 \xW \bigr)^2_{ } 
 \, V^\rmii{LO}_{P;\mu\rho}
 \, \Delta^{-1}_{P;\rho\sigma}
 \, V^\rmii{LO}_{P;\sigma\nu}
 \nn[2mm]
 & + & 
 \, e^2 \bigl( 2\delta^{ }_{a,e} - 1 + 4 \xW \bigr) \, C^{ }_a 
 \, V^\rmii{LO}_{P;\mu\rho}
 \, \Delta^{-1}_{P;\rho\sigma}
 \, (P^{ }_\sigma P^{ }_\nu - P^2_{ }\delta^{ }_{\sigma\nu} ) 
 \;.  \la{Smunu} 
\ea
Here $\Delta^{-1}_{K;\nu^{ }_a} \equiv (i\bsl{K})^{-1}_{ }$
is the tree-level neutrino propagator, 
$
 \Delta^{-1}_{P;\rho\sigma} 
$
is the tree-level photon propagator, 
and
sum-integrals over repeated four-momenta ($P,Q$) are implied. 
The last two lines of \eq\nr{Smunu} include fermion-line 
``disconnected'' contributions, 
originating from the last term in \fig\ref{fig:processes}(b), 
and ``LO'' stands for a leading-order evaluation of \eq\nr{Vmunu}.  

%

Now, particles can to a good approximation be treated as massless
if their mass is less than $\sim \pi T$. If we consider temperatures
$T \gsim 1$~MeV, the electron mass does satisfy this requirement. Therefore, 
the electron mass can be set to zero, simplifying the analysis. 

Let us rewrite \eq\nr{Smunu} in the massless limit. 
The neutrino propagators anticommute with~$\gamma^{ }_5$, 
and in the massless limit the electron propagators do the same. 
The mixed correlator  from \eq\nr{Mmunu} vanishes.  
Taking also into account that the open 
$(1-\gamma^{ }_5)$'s of the second term of \eq\nr{Smunu} can 
be transported into the projectors in 
\eq\nr{Sigma_full}, we thereby obtain 
\ba
 \lim_{m^{ }_e\to 0}
 \Sigma^{ }_{K;\nu^{ }_a} 
 & \supset & 
 \frac{G_\rmii{F}^2}{4} 
 \gamma^{ }_\mu 
 \Delta^{-1}_{K+P;\nu^{ }_a}
 \widetilde{\mathbbm{S}}^{a}_{P;\mu\nu}
 \,  \gamma^{ }_\nu (1-\gamma^{ }_5)  
 \;, \la{Sigma_full_2}
 \\[3mm]
 \widetilde{\mathbbm{S}}^{a}_{P;\mu\nu}
 & = & 
 \bigl( 2\, 
 {\textstyle \sum_b} \bigr)
 \tr \bigl[\, 
   \Delta^{-1}_{P+Q;\nu^{ }_b} \gamma^{ }_\mu 
   \Delta^{-1}_{Q;\nu^{ }_b} \gamma^{ }_\nu 
 \,\bigr] 
  - 4\,
 \gamma^{ }_\nu\, 
   \Delta^{-1}_{P+Q;\nu^{ }_a} \gamma^{ }_\mu\, 
   \Delta^{-1}_{Q;\nu^{ }_a} 
 \nn[2mm] 
 & - & 
 \, \bigl[ \,
   \bigl( 2\delta^{ }_{a,e} - 1 + 4 \xW \bigr)^2_{ }
 + 1 
 \, \bigr] \,
 V^{ }_{P;\mu\nu}
 \nn[2mm]
 & + & 
 \, e^2 \bigl( 2\delta^{ }_{a,e} - 1 + 4 \xW \bigr)^2_{ } 
 \, V^\rmii{LO}_{P;\mu\rho}
 \, \Delta^{-1}_{P;\rho\sigma}
 \, V^\rmii{LO}_{P;\sigma\nu}
 \nn[2mm]
 & + & 
 \, e^2 \bigl( 2\delta^{ }_{a,e} - 1 + 4 \xW \bigr) \, C^{ }_a 
 \, V^\rmii{LO}_{P;\mu\rho}
 \, \Delta^{-1}_{P;\rho\sigma}
 \, ( P^{ }_\sigma P^{ }_\nu - P^2_{ }\delta^{ }_{\sigma\nu} ) 
 \;. \la{Smunu_massless}  
\ea

%
\subsection{Carrying out a Matsubara sum and angular integrals}
\la{ss:matsubara}

Our next goal is to carry out the outermost Matsubara sum-integral
in \eq\nr{Sigma_full_2}, over $P$. 
In order to simplify the notation, 
let us denote the second part of the integrand by 
$\Pi^{ }_{P}$. We write it in a spectral representation, 
\be
 \Pi^{ }_P = 
 \int_{-\infty}^{\infty} \! \frac{{\rm d}p^0_{ }}{\pi} \, 
 \frac{\im \Pi^{ }_\P}{ p^0_{ } - i p^{ }_n}
 \;, \quad
 \im \Pi^{ }_\P \; \equiv \;  
 \im \Pi^{ }_P \big|^{ }_{p^{ }_n\to -i [p^0_{ } + i 0^+_{ }]}
 \;,
\ee
where $p^{ }_n$ is bosonic. Then we are faced with the structure
\ba
 \Sigma^{ }_{K} & = & 
 \Tint{P}
  \frac{\alpha\,(k^{ }_n + p^{ }_n) + \beta\,(\vec{k+p}) }
       { (k^{ }_n + p^{ }_n)^2_{ } + (\vec{k+p})^2_{ } }
 \, \Pi^{ }_{P}
 \nn 
 & = & 
 \int_\vec{p} 
 \int_{-\infty}^{\infty} \! \frac{{\rm d}p^0_{ }}{\pi} \, 
 \, T \!\! \sum_{p^{ }_n , \{ r^{ }_n \} }
 \delta^{ }_{0,k^{ }_n + p^{ }_n - r^{ }_n}
 \frac{\alpha\, r^{ }_n + \beta\,(\vec{k+p})}{r_n^2 + \epsilon^2_{kp}}
 \, 
 \frac{ \im \Pi^{ }_\P }{p^0_{ }- i p^{ }_n}
 \;, 
\ea
where $k^{ }_n$ and $r^{ }_n$ are fermionic, and 
$\epsilon^{2}_{kp} \equiv (\vec{k+p})^2$. 
We may now write 
$
 \delta^{ }_{0,k^{ }_n + p^{ }_n - r^{ }_n} 
 = 
 T \int_0^{1/T} \! {\rm d}\tau \, e^{i(k^{ }_n + p^{ }_n - r^{ }_n)\tau}_{ }
$, 
and carry out the sums over $p^{ }_n$ and $r^{ }_n$.
Furthermore the integral over $\tau$ is doable, yielding
\ba
 \Sigma^{ }_K & = & 
 \int_\vec{p} 
 \int_{-\infty}^{\infty} \! \frac{{\rm d}p^0_{ }}{\pi} \, 
 \frac{ 1 }{2\epsilon^{ }_{kp}}
 \biggl\{\, 
  \bigl[ i \alpha \epsilon^{ }_{kp} - \beta(\vec{k+p})\bigr]
  \frac{\nF^{ }(\epsilon^{ }_{kp}) + \nB^{ }(p^0_{ })}
       {i k^{ }_n + p^0_{ }- \epsilon^{ }_{kp}}
 \nn 
 & & \hspace*{1.5cm}
 + \, 
  \bigl[ i \alpha \epsilon^{ }_{kp} + \beta(\vec{k+p})\bigr]
  \frac{1 - \nF^{ }(\epsilon^{ }_{kp}) + \nB^{ }(p^0_{ })}
       {i k^{ }_n + p^0_{ } + \epsilon^{ }_{kp}}
 \,\biggr\}  \, \im \Pi^{ }_\P
 \;,
\ea
where $\nB^{ }$ and $\nF^{ }$ denote the Bose and Fermi distributions, 
respectively. 

As a next step, we continue to Minkowskian frequency, 
$i k^{ }_n \to k^0_{ } + i 0^+_{ }$, and take the imaginary part. 
We also substitute 
$
 p^0_{ } \to - p^0_{ }
$, 
$
 \vec{p} \to -\vec{p}
$, 
and note that bosonic spectral functions are odd under this substitution. 
Finally, 
$
 \epsilon^{ }_{kp}
$
is redefined as 
$
 \epsilon^{ }_{kp} \equiv |\vec{k-p}|
$, 
and we recall that 
$\nB^{ }(-x) = -1 - \nB^{ }(x)$, 
$\nF^{ }(-x) =  1 - \nF^{ }(x)$.
This yields
\ba
 \im \Sigma^{ }_{\K} & = & 
 \int_\vec{p} 
 \int_{-\infty}^{\infty} \! \frac{{\rm d}p^0_{ }}{2\epsilon^{ }_{kp}}\,
  \bigl[\, i \alpha (p^0_{ } - k^0_{ }) + \beta(\vec{k - p}) \,\bigr]
  \bigl[\, 1 - \nF^{ }(k^0_{ } - p^0_{ }) + \nB^{ }(p^0_{ }) \,\bigr]
 \nn[3mm] 
 & \times & 
  \bigl[\,  
            \delta(k^0_{ }- p^0_{ } - \epsilon^{ }_{kp})
          - 
            \delta(k^0_{ }- p^0_{ } + \epsilon^{ }_{kp})   
   \,\bigr]
 \, \im \Pi^{ }_\P
 \;. \la{rho_1}
\ea

Subsequently, we may integrate over the angles of $\vec{p}$, 
removing the Dirac-$\delta$'s. At this point we also 
go to the light cone, $k^0_{ } = k$, relevant for on-shell neutrinos. 
The first Dirac-$\delta$ can be seen to be realized for 
$|p^0_{ }| < p$, i.e.\ it corresponds to a $t$-channel process. 
The second Dirac-$\delta$ is realized for $p^0_{ } > p$, 
and corresponds to an $s$-channel process. For both channels, 
the Dirac-$\delta$'s imply
\be
 \vec{k}\cdot\vec{p} 
 = \frac{k^2 + p^2 - \epsilon^2_{kp}}{2}
 \; \stackrel{k^0_{ } = k}{=} \; 
 k p^0_{ } - \frac{\P^2_{ }}{2}
 \;. \la{angles} 
\ee
Introducing
\be
 p^{ }_{\pm} \; \equiv \; \frac{p^0_{ } \pm p}{2}
 \;, \la{ppm}
\ee
\eq\nr{rho_1} can be turned into
\ba
 \im \Sigma^{ }_{k} & = & 
 \frac{1}{4\pi^2 k}
 \biggl(\,
 \overbrace{
  \int_{-\infty}^0 \! {\rm d} p^{ }_{-}  
  \int_0^k \! {\rm d} p^{ }_{+} 
 }^{\rm {\it t}-channel}
 -   
 \overbrace{
  \int_0^k \! {\rm d} p^{ }_{-} 
  \int_k^\infty \! {\rm d} p^{ }_+
  }^{\rm {\it s}-channel}
 \,\biggr)
 \nn[3mm] & \times & 
 \, p \, 
  \bigl[\, i \alpha (p^0_{ } - k ) + \beta(\vec{k - p}) \,\bigr]
  \bigl[\, 1 - \nF^{ }(k - p^0_{ }) + \nB^{ }(p^0_{ }) \,\bigr]
 \, \im \Pi^{ }_\P
 \;. \hspace*{4mm} \la{rho_2}
\ea

In this section we have dealt with the outer sum-integral
in \eq\nr{Sigma_full_2}. 
However, the same procedure can also be used for 
the inner one, except that then the four-momentum $\P$ is 
not light-like. In some cases, it also turns out to be 
helpful to redefine variables, whereby a clean 
separation into an outer and inner sum-integration
no longer applies (cf.\ appendix~\ref{ss:lo}).

%
\subsection{Projecting out the interaction rate}
\la{ss:proj}

The object in \eq\nr{rho_2} is a matrix in Dirac space, 
cf.\ \eq\nr{Sigma_full_2}. Assuming that all the sum-integrals have been 
carried out, going over to Minkowskian signature
(whereby $i\bsl{K}\to \bsl{\K}$), 
and redundantly displaying chiral projectors in accordance
with \eq\nr{prop_full},  
the result has the form~\cite{weldon} 
\be
 \Sigma^{ }_{k;\nu^{ }_a} 
 \; = \; 
 \aR\, \bigl(\, a \bsl{\K} + b \msl{u} \,\bigr)\, \aL
 \;, \la{b_def}
\ee
where $u \equiv (1,\vec{0})$ denotes the medium four-velocity. 

The interaction rate influences the time evolution of on-shell 
active neutrinos (cf.\ \eq\nr{vn_1}).
In terms of \eq\nr{b_def}, 
it can be found in the imaginary part of $b$, 
\be
 b \; = \; b^{ }_r + \frac{i}{2} \, 
 \Gamma^{ }_{k;\nu^{ }_a}
 \;, \quad
 b^{ }_r \in \mathbbm{R}
 \;. \la{b_full}
\ee
The real part $b^{ }_r$, 
which gives a medium modification to the effective Hamiltonian
$\Delta\mathcal{E}^{ }_k$ appearing in \eq\nr{vn_1}, 
is of $\rmO(G^{ }_\rmii{F})$~\cite{raffelt}. 
The imaginary part, which is of $\rmO(G^{2}_\rmii{F})$
and interests us here, 
can be projected out as 
\be
 \Gamma^{ }_{k;\nu^{ }_a} 
 = \frac{1}{k} \im \tr [ \bsl{\K}
  \Sigma^{ }_{k;\nu^{ }_a} 
  ]
 \;. \la{heli_proj}
\ee

%

As a next step, we put together the ingredients from 
\eqs\nr{Sigma_full_2}, \nr{Smunu_massless}, 
\nr{rho_2} and \nr{heli_proj}. 
First we consider
the terms from \eq\nr{Smunu_massless} that have 
a factorized Dirac trace, as is indeed the case in 
all but one term. We then find
\ba
 \Gamma^{ }_{k;\nu^{ }_a}
 & \supset & 
 \im\biggl\{\, 
 \frac{G_\rmii{F}^2}{4k} 
 \frac{ \tr\bigl[\, 
 ( i \bsl{K} ) \gamma^{ }_\mu 
 (-i)( \bsl{K} + \bsl{P} ) \gamma^{ }_\nu
 (1-\gamma^{ }_5)
 \,\bigr]
 }{(K+P)^2_{ }} 
 \,
 \widetilde{\mathbbm{S}}^{a}_{P;\mu\nu}
 \,\biggr\}
 \nn 
 & \to & 
 \im\biggl\{\, 
 \frac{G_\rmii{F}^2}{k}
 \frac{2 K^{ }_\mu K^{ }_\nu - \delta^{ }_{\mu\nu}K\cdot(K+P)}
 {(K+P)^2} 
 \,
 \widetilde{\mathbbm{S}}^{a}_{P;\mu\nu}
 \,\biggr\}
 \;,
\ea
where we noted that the part involving $\gamma^{ }_5$ drops out, 
if $ \widetilde{\mathbbm{S}}^{a}_{P;\mu\nu} $ is symmetric in 
$\mu \leftrightarrow \nu$, and that the parts proportional to 
$P^{ }_\mu$ and $P^{ }_\nu$ drop out, if 
$ \widetilde{\mathbbm{S}}^{a}_{P;\mu\nu} $ is transverse.

Subsequently, we make use of \eq\nr{rho_2}, with 
$ \widetilde{\mathbbm{S}}^{a}_{P;\mu\nu} $ playing the 
role of $\Pi^{ }_P$. Inserting \eq\nr{angles}, this leads to 
\ba
 \Gamma^{ }_{k;\nu^{ }_a}
 & \supset & 
 \frac{G_\rmii{F}^2}{8\pi^2k^2} 
 \biggl(\,
 \overbrace{
  \int_{-\infty}^0 \! {\rm d} p^{ }_{-}  
  \int_0^k \! {\rm d} p^{ }_{+} 
 }^{\rm {\it t}-channel}
 -   
 \overbrace{
  \int_0^k \! {\rm d} p^{ }_{-} 
  \int_k^\infty \! {\rm d} p^{ }_+
  }^{\rm {\it s}-channel}
 \,\biggr)
 \nn[3mm] & \times & 
 \, p \, 
  \bigl[\, 4 K^{ }_\mu K^{ }_\nu - \delta^{ }_{\mu\nu} \P^2 \,\bigr]
  \bigl[\, 1 - \nF^{ }(k - p^0_{ }) + \nB^{ }(p^0_{ }) \,\bigr]
 \, \im\widetilde{\mathbbm{S}}^{a}_{\P;\mu\nu}
 \;. \la{rho_3}
\ea
To complete the transition to Minkowskian spacetime, we recall
the rules from \eq\nr{rules}, resulting in
\ba
 \Gamma^{ }_{k;\nu^{ }_a}
 & \supset & 
 \frac{G_\rmii{F}^2}{8\pi^2k^2} 
 \biggl(\,
 \overbrace{
  \int_{-\infty}^0 \! {\rm d} p^{ }_{-}  
  \int_0^k \! {\rm d} p^{ }_{+} 
 }^{\rm {\it t}-channel}
 -   
 \overbrace{
  \int_0^k \! {\rm d} p^{ }_{-} 
  \int_k^\infty \! {\rm d} p^{ }_+
  }^{\rm {\it s}-channel}
 \,\biggr)
 \nn[3mm] & \times & 
 \, p \, 
  \bigl[\, - 4 \K_\mu^{ } \K^{ }_\nu -
    \eta_{\mu\nu}^{ } \P^2 \,\bigr]
  \bigl[\, 1 - \nF^{ }(k - p^0_{ }) + \nB^{ }(p^0_{ }) \,\bigr]
 \, \im\widetilde{\mathbbm{S}}^{a;\mu\nu}_{\P} 
 \;. \la{rho_4}
\ea

As a final ingredient, we take into account that 
the spectral functions originating from the factorized terms
in \eq\nr{Smunu_massless} are transverse. Anticipating 
\se\ref{se:nlo}, let us focus on  
\be
  \im \widetilde{\mathbbm{S}}^{a;\mu\nu}_{\P}
  \supset
 - 
 \, \bigl[ \,
   \bigl( 2\delta^{ }_{a,e} - 1 + 4 \xW \bigr)^2_{ }
 + 1 
 \, \bigr] \,
 \im V^{\mu\nu}_{\P}
 \;. \la{e_part}
\ee
The vector channel spectral function can be decomposed as 
\be
 \im V^{\mu\nu}_{\P} = 
 \mathbbm{P}^{\mu\nu}_\rmii{T} \, \rho^{ }_\rmii{T} + 
 \mathbbm{P}^{\mu\nu}_\rmii{L} \, \rho^{ }_\rmii{L}
 \;, \la{decomposition}
\ee
where the projectors have been defined as 
\be
 \mathbbm{P}^{\mu\nu}_\rmii{T} \; \equiv \; 
 - \eta^{\mu}_{i}\eta^{\nu}_j
   \,\biggl( \delta^{ }_{ij} - \frac{p^{ }_i p^{ }_j}{p^2} \biggr)
 \;, \quad
 \mathbbm{P}^{\mu\nu}_\rmii{L} \; \equiv \; 
 \eta^{\mu\nu}_{ } 
 - \frac{\P^\mu_{ } \P^\nu_{ }}{\P^2_{ }}
 - \mathbbm{P}^{\mu\nu}_\rmii{T}
 \;.
\ee
Inserting this representation, and making use of 
\eq\nr{angles} in order to resolve the scalar
products $\vec{k}\cdot\vec{p}$
originating from 
$
 \K^{ }_\mu \K^{ }_\nu \mathbbm{P}^{\mu\nu}_\rmii{T}
$, the contraction in \eq\nr{rho_4}
takes the form
\ba
 && \hspace*{-1.5cm} 
 \bigl[\, - 4 \K^{ }_\mu \K^{ }_\nu - \eta^{ }_{\mu\nu} \P^2 \,\bigr]
 \im V^{\mu\nu}_{\P}
 \nn[2mm] 
  & = & 
 - \frac{\P^2_{ }}{2}
 \biggl\{
   \bigl( 2\rho^{ }_\rmii{T} + \rho^{ }_\rmii{L} \bigr)  
   \biggl[
     1 + \biggl( \frac{2k - p^0_{ }}{p} \biggr)^2_{ } 
   \,\biggr]
   + 
   \rho^{ }_\rmii{L} \, 
   \biggl[
     1  - 3\, \biggl( \frac{2k - p^0_{ }}{p} \biggr)^2_{ } 
   \,\biggr]
 \biggr\}
 \;. \la{rho_qed}
\ea
This representation will be employed in \se\ref{se:nlo}, 
and can also be crosschecked numerically, as demonstrated
in \se\ref{ss:numerics}.

%
\subsection{HTL resummation of soft $t$-channel photons}
\la{ss:htl}

The computation described above employs
tree-level propagators. While the result is UV finite, 
one of the integrals
turns out to be logarithmically IR divergent. 
In the language of \eq\nr{rho_4}, 
the IR divergence emerges from the $t$-channel contribution, 
when the momentum flowing through the photon propagator, 
$
 \Delta^{-1}_{P;\rho\sigma}
$
in \eq\nr{Smunu_massless}, 
is ``soft'', 
$| p^0_{ } | \le p \ll \pi T$.
The technical reason is that the vector correlators turn
into ``Hard Thermal Loops'' (HTLs) in this domain,  
$V^\rmii{LO}_{P} \sim T^2$. Then they do not regulate that IR regime
any more, as would be the case in vacuum, where they 
behave as $\sim P^2$. In addition, soft photons are 
Bose-enhanced, with 
$
 \nB^{ }(p^0_{ }) \approx T/p^0_{ }
$.

HTLs appear not only in 
the vertices $V^\rmii{LO}_{P}$,
but they also influence the self-energies of the photons. 
When all effects scaling as $\sim T^2$ are included, we talk about
HTL resummation~\cite{htl3,htl4}. HTL resummation introduces screening
effects (Debye screening, Landau damping) to the photon propagator, 
and these regulate the regime of soft momentum exchange. 

In our case, when only one contribution is affected, 
HTL resummation is fairly simple to implement. The relevant 
structures originate from 
the IR limits of the real and imaginary parts of $V^\rmii{LO}_{\P}$.
The imaginary parts are obtained from \eqs\nr{rhoV} and \nr{rho00}, 
\ba
 \rho_\rmii{T}^\rmii{IR}
 & \equiv & 
 [\, \rho_\rmii{T}^\rmii{LO} \,]^{ }_{p,|p^0_{ }| \ll \pi T}
 \; = \; 
 + \frac{\pi p^0_{ }\P^2_{ } T^2_{ }}{12 p^3_{ }}
 \,\theta(p - |p^0_{ }|)
 \;, \la{rhoTir} \\[2mm]
 \rho_\rmii{L}^\rmii{IR}
 & \equiv & 
 [\, \rho_\rmii{L}^\rmii{LO} \,]^{ }_{p,|p^0_{ }| \ll \pi T}
 \; = \; -
 \frac{\pi p^0_{ }\P^2_{ } T^2_{ }}{6 p^3_{ }}
 \,\theta(p - |p^0_{ }|)
 \;. \la{rhoLir}
\ea
The real parts can be derived from \eqs\nr{chiV} and \nr{chi00}, 
\ba
 \chi_\rmii{T}^\rmii{IR}
 & \equiv & 
 [\, \chi_\rmii{T}^\rmii{LO} \,]^{ }_{p,|p^0_{ }| \ll \pi T}
 \; = \; + 
 \frac{T^2_{ }}{6 p^2_{ }}
 \biggl[\,  
   (p^0_{ })^2_{ }
 - \frac{p^0_{ } \P^2_{ }}{2 p}
   \ln\biggl| \frac{p+p^0_{ }}{p-p^0_{ }} \biggr|
 \,\biggr]
 \;, \la{chiTir} \\[2mm] 
 \chi_\rmii{L}^\rmii{IR}
 & \equiv & 
 [\, \chi_\rmii{L}^\rmii{LO} \,]^{ }_{p,|p^0_{ }| \ll \pi T}
 \; = \; - 
 \frac{\P^2_{ }T^2_{ }}{3 p^2_{ }}
 \biggl[\,  
 1 
 - \frac{p^0_{ }}{2 p}
   \ln\biggl| \frac{p+p^0_{ }}{p-p^0_{ }} \biggr|
 \,\biggr]
 \;. \la{chiLir}
\ea

With the help of \eqs\nr{rhoTir}--\nr{chiLir}, we can 
construct the resummed photon propagator, 
denoted by $\Delta^{-1*}_{P;\mu\nu}$. 
After analytic continuation, its transverse part can be 
decomposed in analogy with \eq\nr{decomposition}
(the longitudinal part depends on the gauge parameter and 
drops out when inserted into physical observables). For convenience
we insert an overall minus sign, and denote the 
real and imaginary parts of the 
resummed propagator by 
$\mathcal{R}^*_\rmii{T,L}$ and $\mathcal{I}^*_\rmii{T,L}$, 
\ba
  - \Delta^{-1*;\mu\nu}_{\P}
 & \equiv & 
 \mathbbm{P}^{\mu\nu}_\rmii{T} \, 
  \bigl( \mathcal{R}^{*}_\rmii{T} + i \mathcal{I}^{*}_\rmii{T} \bigr) 
 + 
 \mathbbm{P}^{\mu\nu}_\rmii{L} \, 
  \bigl( \mathcal{R}^{*}_\rmii{L} + i \mathcal{I}^{*}_\rmii{L} \bigr) 
 \;.
\ea
With the help of \eqs\nr{rhoTir}--\nr{chiLir}, we find
\ba
 \mathcal{R}^*_\rmii{T,L} 
 & = &  
 \re \biggl\{ \frac{1}{\P^2_{ } - e^2_{ } (\chi^\rmiii{IR}_\rmiii{T,L} +
 i \rho^\rmiii{IR}_\rmiii{T,L} )}
 \biggr\}
 \; = \; 
 \frac{\P^2_{ } - e^2_{ } \chi^\rmiii{IR}_\rmiii{T,L}}
 {(\P^2_{ } - e^2_{ } \chi^\rmiii{IR}_\rmiii{T,L})^2_{ }
 + (e^2_{ } \rho^\rmiii{IR}_\rmiii{T,L})^2_{ }
 } 
 \;, \la{R}
\ea
and correspondingly for the imaginary parts. 

In order to implement the resummation in an unambiguous way, 
we only employ it in terms where it is necessary at $\rmO(e^2)$, 
but not in terms where its effect is of higher order in $e$. 
Notably, in the $t$-channel, we only keep the contribution from 
$\mathcal{R}^*_\rmii{T,L}$, but not from $\mathcal{I}^*_\rmii{T,L}$, 
since the latter leads to a contribution of $\rmO(e^4)$.\footnote{%
 In terms of \fig\ref{fig:processes}, this corresponds to 
 the square of the $t$-channel amplitude in set~(g), rather than its 
 interference with the tree-level process.
 }
In the $s$-channel, we set
$\mathcal{R}^*_\rmii{T,L} \to \mathcal{R}^{ }_\rmii{T,L} \equiv 1/\P^2_{ }$, 
since the integral is IR-finite. Furthermore, we omit the effect
from $\mathcal{I}^*_\rmii{T,L}$ in the $s$-channel as well.
It would lead to the processes 
illustrated in set~(h) of \fig\ref{fig:processes}. 
These correspond to the famous 
plasmon contributions~(cf.,\ e.g.,\ ref.~\cite{old}). 
However, they are formally of $\rmO(e^3)$ rather than $\rmO(e^2)$:
apart from the vertices, the result of a $1\leftrightarrow 2$ 
reaction is phase-space suppressed by the plasmon mass $\sim eT$.

After the inclusion of HTL resummation in the $t$-channel, 
the contribution from soft momentum exchange is finite. 
However, it is logaritmically enhanced, by $\ln(1/e^2_{ })$. 
The coefficient of this logarithm can be
worked out analytically, as shown in appendix~B. The final result, 
formally the largest NLO QED correction that we find, is 
displayed in \eq\nr{Dt_log}. 

%
\section{NLO computation and result}
\la{se:nlo}

%
\begin{figure}[p]
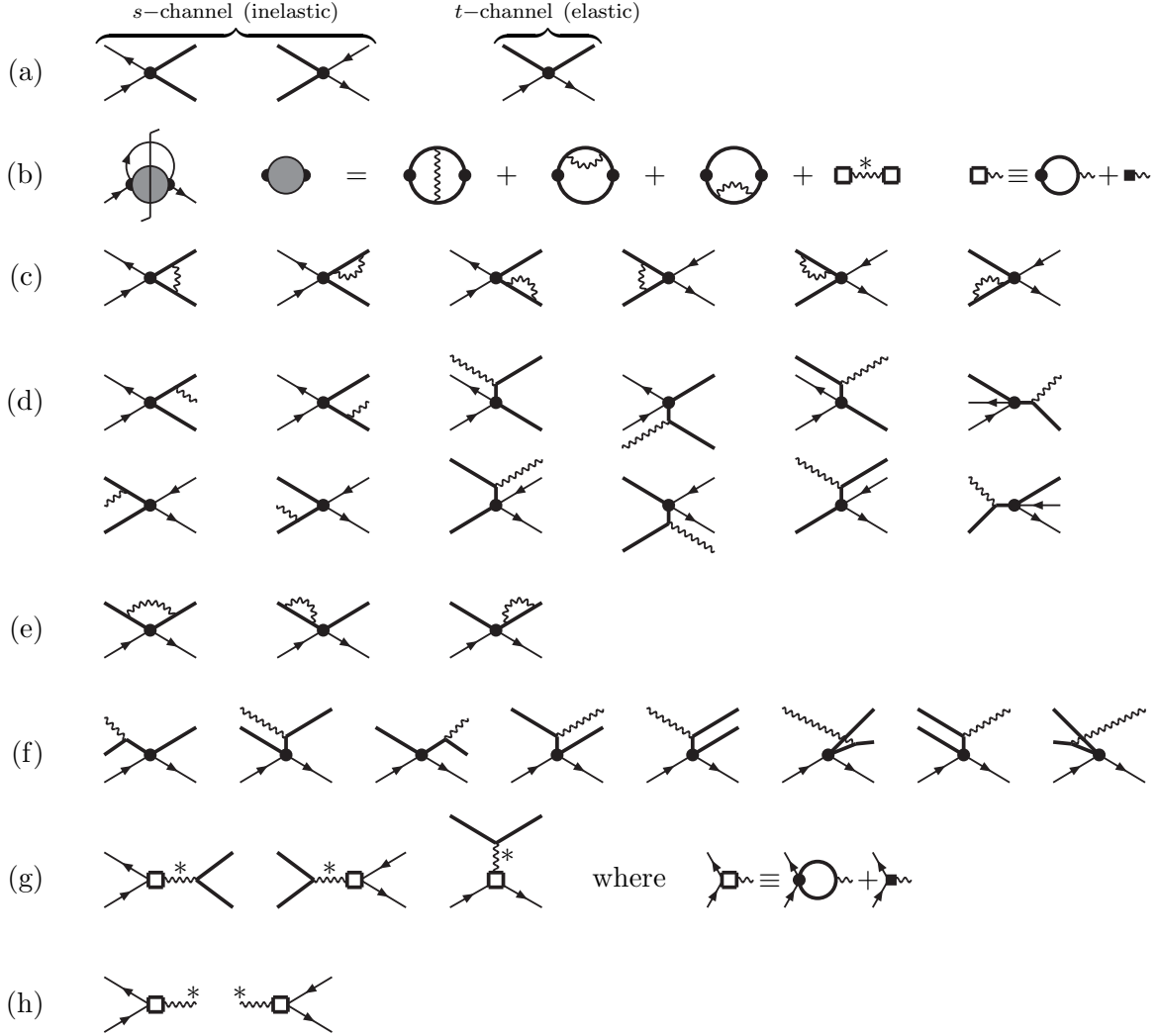


\begin{eqnarray}
\mbox{(a)}
&&
 \hspace*{0mm}
 \overbrace{ 
 \ampLOsa \hspace*{10mm} 
 \ampLOsb \; }^{s{\rm -channel~(inelastic)}}
 \hspace*{10mm}
 \overbrace{ 
 \ampLOt  \; }^{t{\rm -channel~(elastic)}}
 \nn[5mm]
\mbox{(b)}
&&
 \hspace*{0mm}
 \cutNLO  \hspace*{5mm}
 \NLOblob \hspace*{1mm} = \hspace*{1mm}
 \selfEa  \hspace*{1mm} + \hspace*{1mm}
 \selfEb  \hspace*{1mm} + \hspace*{1mm}
 \selfEc  \hspace*{1mm} + \hspace*{-1mm}
 \selfEd  \hspace*{5mm}
 \selfEf  
 \nn[5mm]
\mbox{(c)}
&&
 \hspace*{0mm}
 \virtNLOsa \hspace*{10mm} 
 \virtNLOsb \hspace*{10mm}
 \virtNLOsc \hspace*{10mm}
 \virtNLOsd \hspace*{10mm}
 \virtNLOse \hspace*{10mm}
 \virtNLOsf \hspace*{10mm}
 \nn[8mm]
\mbox{(d)}
&&
 \hspace*{0mm}
 \realNLOsa \hspace*{10mm} 
 \realNLOsb \hspace*{10mm} 
 \realNLOsc \hspace*{10mm} 
 \realNLOsd \hspace*{10mm} 
 \realNLOse \hspace*{10mm} 
 \realNLOsh \hspace*{10mm} 
 \nn[5mm]
&&
 \hspace*{0mm}
 \realNLOsi \hspace*{10mm} 
 \realNLOsj \hspace*{10mm} 
 \realNLOsk \hspace*{10mm} 
 \realNLOsl \hspace*{10mm} 
 \realNLOsm \hspace*{10mm} 
 \realNLOsp \hspace*{10mm} 
 \nn[8mm]
\mbox{(e)}
&&
 \hspace*{0mm}
 \virtNLOta \hspace*{10mm} 
 \virtNLOtb \hspace*{10mm}
 \virtNLOtc \hspace*{10mm}
 \nn[8mm]
\mbox{(f)}
&&
 \hspace*{0mm}
 \realNLOta \hspace*{5mm} 
 \realNLOte \hspace*{5mm} 
 \realNLOtb \hspace*{5mm} 
 \realNLOtc \hspace*{5mm} 
 \realNLOtd \hspace*{5mm} 
 \realNLOtg \hspace*{5mm} 
 \realNLOtf \hspace*{5mm} 
 \realNLOth \hspace*{5mm} 
 \nn[8mm]
\mbox{(g)}
&&
 \hspace*{0mm}
 \extrNLOc \hspace*{10mm} 
 \extrNLOd \hspace*{10mm} 
 \extrNLOe \hspace*{7mm}
 \mbox{where} \hspace*{5mm}
 \selfEfX  
 \nn[8mm]
\mbox{(h)}
&&
 \hspace*{0mm}
 \extrNLOa \hspace*{5mm} 
 \extrNLOb 
 \nonumber
\end{eqnarray}

\vspace*{1mm}

\caption[a]{\small 
 (a)~leading-order reactions between neutrinos (arrowed lines)
 and electrons and positrons (thick lines, with arrows 
 omitted, understanding that both directions should be included);
 (b)~a dispersive representation of the NLO corrections, with the 
 vertical line denoting a cut, a wiggly line a photon,
 an asterisk HTL-resummation,  
 and the filled box the loop-generated operator from \eq\nr{ct_1};
 (c)~virtual corrections to $s$-channel reactions;
 (d)~real corrections to $s$-channel reactions;
 (e)~virtual corrections to $t$-channel reactions; 
 (f)~real corrections to $t$-channel reactions; 
 (g)~$s$ and $t$-channel processes involving virtual photon 
 exchange, originating from the last term in set~(b); 
 (h)~$s$-channel processes involving a plasmon resonance, 
 originating from the last term in set~(b). 
} 
\la{fig:processes}
\end{figure}
%

We now put together the full NLO expression for the thermal interaction
rate of a neutrino of flavour $a$ and momentum $k$. 
The result is expressed in terms of 
partial interaction rates, in the spirit of \eq\nr{vn_2}. 
We make use of 
the Weinberg angle parametrization from \eq\nr{xW}, 
the simplified integrals in \eqs\nr{Xit} and \nr{Xis}
for scatterings off neutrinos, 
the full representation of the $e^+_{ }e^-_{ }$ contribution 
from \eqs\nr{rho_4}, \nr{e_part} and \nr{rho_qed}, 
the two independent electron-positron spectral functions defined 
according to \eq\nr{decomposition}, 
as well as HTL resummation from \eq\nr{R}. 
We assume all ensembles to carry the same 
temperature, and omit chemical potentials. 
Then scatterings off, pair annihilations into, or pair creations from,  
the same neutrino flavour, different neutrino flavours, 
and electrons and positrons, yield the contributions
\ba
 && \hspace*{-0.8cm}
 \Gamma^{(\nu^{ }_a)}_{k;\nu^{ }_a} 
 \; = \; 
 \Bigl\{ 2 \, + 2  \,  
 \Bigr\} \,  
 \frac{G_\rmii{F}^2}{16\pi^3 k^2}
 \bigl[ \, 
  2 \Xi^t_{ }+ \Xi^s_{ }
 \, \bigr]
 \;, \la{rho_8a} \\[3mm]
 && \hspace*{-0.8cm} 
 \Gamma^{(\nu^{ }_b \neq \nu^{ }_a)}_{k;\nu^{ }_a} 
 \; = \; 
 \Bigl\{ (2 {\textstyle \sum_{b\neq a}} )
 \Bigr\} \,  
 \frac{G_\rmii{F}^2}{16\pi^3 k^2}
 \bigl[ \, 
  2 \Xi^t_{ }+ \Xi^s_{ }
 \, \bigr]
 \;, \la{rho_8b} \\[3mm]
 && \hspace*{-0.8cm}
 \Gamma^{(e^+_{ }e^-_{ })}_{k;\nu^{ }_a}
 \; = \; 
 \frac{G_\rmii{F}^2}{8\pi^2k^2} 
 \biggl(\,
 \overbrace{
  \int_{-\infty}^0 \! {\rm d} p^{ }_{-}  
  \int_0^k \! {\rm d} p^{ }_{+} 
 }^{\rm {\it t}-channel}
 -   
 \overbrace{
  \int_0^k \! {\rm d} p^{ }_{-} 
  \int_k^\infty \! {\rm d} p^{ }_+
  }^{\rm {\it s}-channel}
 \,\biggr)
 \, 
  {p}\, \P^2_{ }
 \, 
  \bigl[\, 1 - \nF^{ }(k - p^0_{ }) + \nB^{ }(p^0_{ }) \,\bigr]
 \nn[2mm] 
  & \times & 
 \hspace*{-3mm} 
 \biggl\{ 
  \, \bigl[ \,
   \bigl( 2\delta^{ }_{a,e} - 1 + 4 \xW \bigr)^2_{ }
 + 1 
 \, \bigr] \,
 \biggl[
   \rho^\rmii{NLO}_\rmii{T} + \rho^\rmii{NLO}_\rmii{L} 
  + 
   \bigl( \rho^\rmii{NLO}_\rmii{T} - \rho^\rmii{NLO}_\rmii{L} \bigr)  
   \biggl( \frac{2k - p^0_{ }}{p} \biggr)^2_{ } 
 \biggr]
 \nn[2mm]
 & + & 
 \hspace*{-3mm} 
 2 e^2 \bigl( 2\delta^{ }_{a,e} - 1 + 4 \xW \bigr)^2_{ }
 \biggl[
    \rho^\rmii{LO}_\rmii{T} \chi^\rmii{LO}_\rmii{T} \mathcal{R}^*_\rmii{T}
  + \rho^\rmii{LO}_\rmii{L} \chi^\rmii{LO}_\rmii{L} \mathcal{R}^*_\rmii{L}
  + 
   \bigl(
    \rho^\rmii{LO}_\rmii{T} \chi^\rmii{LO}_\rmii{T} \mathcal{R}^*_\rmii{T}
  - \rho^\rmii{LO}_\rmii{L} \chi^\rmii{LO}_\rmii{L} \mathcal{R}^*_\rmii{L}
   \bigr)  
   \biggl( \frac{2k - p^0_{ }}{p} \biggr)^2_{ } 
 \biggr]
 \nn[2mm]
 & + & 
 \hspace*{-3mm} 
 e^2 C^{ }_a \bigl( 2\delta^{ }_{a,e} - 1 + 4 \xW \bigr)^{ }_{ } 
 \,
 \biggl[
    \rho^\rmii{LO}_\rmii{T} 
  + \rho^\rmii{LO}_\rmii{L} 
  + 
   \bigl( \rho^\rmii{LO}_\rmii{T} 
        - \rho^\rmii{LO}_\rmii{L} 
   \bigr)  
   \biggl( \frac{2k - p^0_{ }}{p} \biggr)^2_{ } 
 \biggr]
 \biggr\}
 \;, \la{rho_8c}
\ea
respectively, where $p^{ }_{\pm}$ are defined according to \eq\nr{ppm}; 
$\P^2 \equiv (p^0_{ })^2_{ } - p^2 = 4 p^{ }_+ p^{ }_-$;
and $\nF^{ }$ and $\nB^{ }$ denote the Fermi and Bose 
distribution functions. The functions $\rho^\rmii{LO,NLO}_\rmii{T,L}$
originate from the imaginary part of the vector current correlator
in \eq\nr{Vmunu}, and are discussed below; 
the functions $\chi^\rmii{LO}_\rmii{T,L}$
originate from its real part,  
and are given in appendix~\ref{ss:reV}.

Let us elaborate on the three contributions
in \eq\nr{rho_8c}. The first line contains the 
NLO spectral functions $\rho^\rmii{NLO}_\rmii{T,L}$.
These represent the diagrams in 
\fig\ref{fig:processes}(c)--(f):
all of these processes,  
and the interference terms between them,
are contained in the evaluations of 
$\rho^\rmii{NLO}_\rmii{T,L}$ that were carried out in 
refs.~\cite{twoloop,twoloop_code}. We just need to adjust 
the QCD coupling $g^2_{ }$ and the Casimir factor $\CF^{ }$
as $g^2_{ }\CF^{ }\to e^2 \equiv 4\pi \alpha^{ }_\rmi{em}$.
We remark in passing that both spectral functions are 
independent of the renormalization scale at NLO. 

The second line of \eq\nr{rho_8c}
originates from the processes
in \fig\ref{fig:processes}(g), once these interfere  
with LO contributions. 
In the notation
of the last term of \fig\ref{fig:processes}(b), 
the open box is replaced by a closed loop here;  
it is the closed
loop which produces the functions $\chi^\rmii{LO}_\rmii{T,L}$.
We note that the $t$-channel part of 
these processes is IR divergent in naive perturbation
theory, and therefore formally the single largest NLO contribution. 
After HTL resummation through~$\mathcal{R}^*_\rmii{T,L}$, 
as explained in \se\ref{ss:htl}, 
it develops a logarithmic dependence on $e^2_{ }$, 
cf.\ \eq\nr{Dt_log}. 

The third line of \eq\nr{rho_8c} 
also originates from the processes
in \fig\ref{fig:processes}(g), but now the open box is replaced
by the effective vertex from \eq\nr{ct_1}, parametrized by the
coefficient $C^{ }_a$. This cancels the UV divergences of 
$\chi^\rmii{LO}_\rmii{T,L}$,
but also introduces a finite contribution, cf.\ \eq\nr{C_a}. 
Similarly to photon vacuum polarization, 
the coefficient $C^{ }_a$ is non-perturbative, 
as it is affected by a low-energy hadronic contribution
from the $Z\gamma$ bubble, cf.\ \se\ref{ss:fermi}. 

Apart from the cancellation of the divergence, the concrete 
role of $C^{ }_a$ is to replace the renormalization scale $\bmu$
in \eq\nr{chi_vac} through the physical scale $m^{ }_e$. 
If we represent momenta as $k/T$, our results thereby obtain
a logarithmic sensitivity on $m^{ }_e/T$.

\vspace*{3mm}

For a numerical illustration, we rewrite \eq\nr{rho_8c} as 
\ba
 \frac{ 
  \Gamma^{(e^+_{ }e^-_{ })}_{k;\nu^{ }_a}
 }{ G_\rmii{F}^2 T^4 k / (8\pi^2_{ })}
 & = & 
 \bigl[ \,
   \bigl( 2\delta^{ }_{a,e} - 1 + 4 \xW \bigr)^2_{ }
 + 1 
 \, \bigr] \,
 \Bigl\{\, 
   \underbrace{A}_{\rm from~\rho^\rmiii{LO}_\rmiii{T,L}} 
   + 
   \underbrace{
   \;
   e^2 B 
   \; 
   }_{\rm from~\rho^\rmiii{NLO}_\rmiii{T,L}} 
 \,\Bigr\} 
 \nn[2mm] 
 & + & 
 e^2_{ } \, \bigl( 2\delta^{ }_{a,e} - 1 + 4 \xW \bigr)^2_{ } \, 
 \biggl\{ 
   \frac{1}{6\pi^2}    
   \biggl[
     \underbrace{C 
        - \biggl( 2\ln\frac{m^{ }_e}{T} + \frac{5}{3} \biggr)\, A
     }_{\rm from~{\it C^{ }_a}~and~\chi^\rmiii{LO}_\rmiii{T,L}
        |^{ }_\rmiii{vac}} 
   \biggr]
   + 
   \underbrace{D}_
  {\rm from~\chi^\rmiii{LO}_\rmiii{T,L} |^{ }_\rmiii{\it T}
       \mathcal{R}^*_\rmiii{T,L} } 
 \biggr\} 
 \nn[2mm] 
 & + & 
 e^2_{ } \, \bigl( 2\delta^{ }_{a,e} - 1 + 4 \xW \bigr)^{ }_{ } \, 
 \Bigl\{ 
     \underbrace{8\, (-0.01 \, ... \, 0.01) \,A 
     }_{\rm from~uncertainty~of~{\it C^{ }_a}} 
 \Bigr\} 
 \;. \la{ABCD}
\ea
The couplings are evaluated at the scale $\bmu = m^{ }_e$ as specified
above \eq\nr{C_a}.
The representation is reliable as long as the logarithm $\ln(m^{ }_e/T)$
is of order unity.  
The coefficients can be decomposed into $t$ and $s$-channel contributions, 
$A = A^{(t)}_{ } + A^{(s)}_{}$. Furthermore, $D^{(t)}_{ }$ is logarithmically
sensitive to the coupling, and we express it as
(cf.\ appendix~B)
\be
 D^{(t)}_{ } = 
 D^{(t)}_{1} \ln \biggl( \frac{1}{e^2_{ }} \biggr)
 + 
 D^{(t)}_{2} 
 \;, \quad
 D^{(t)}_{1} \; = \; -\frac{\pi T}{9 k}
 \;. \la{Dt_log}
\ee
The resulting values are illustrated numerically in
appendix~C.
Including the flavour factors, and the 
uncertainty from our estimate for $C^{ }_a$,  
the final results are shown in \fig\ref{fig:results}.

%
\begin{figure}[p]

\centerline{
     \epsfysize=7.0cm\epsfbox{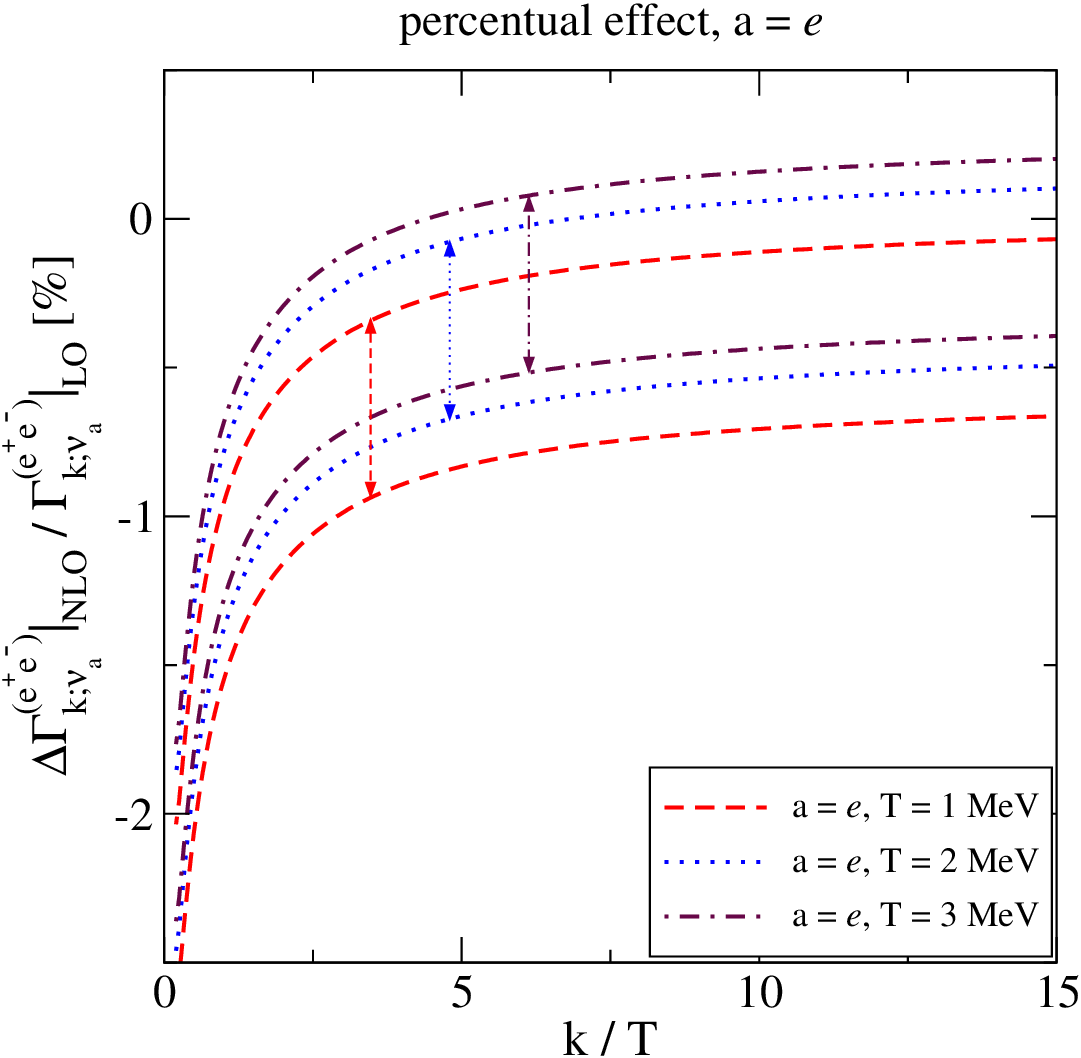}
  ~~~\epsfysize=7.0cm\epsfbox{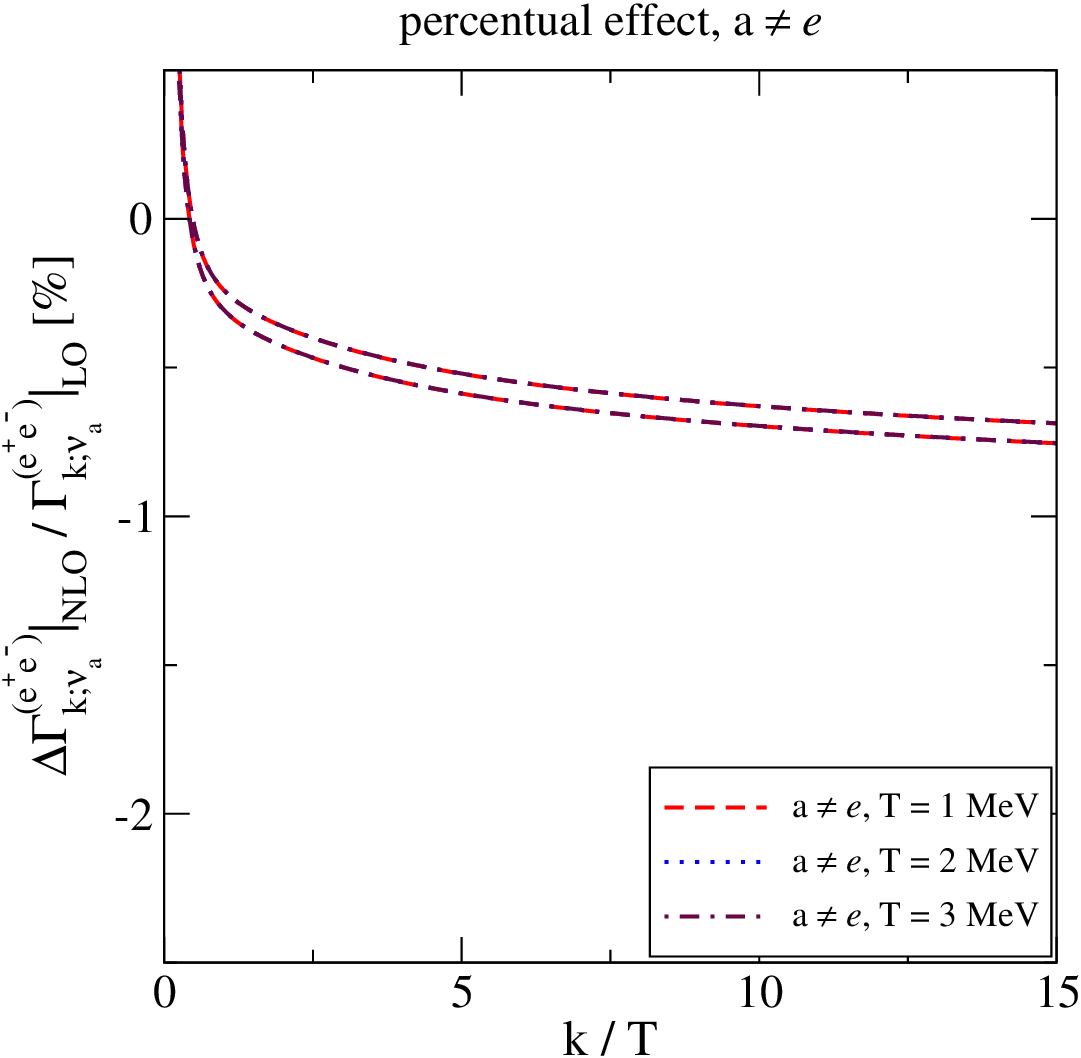}
}

\vspace*{9mm}

\centerline{
     \epsfysize=7.0cm\epsfbox{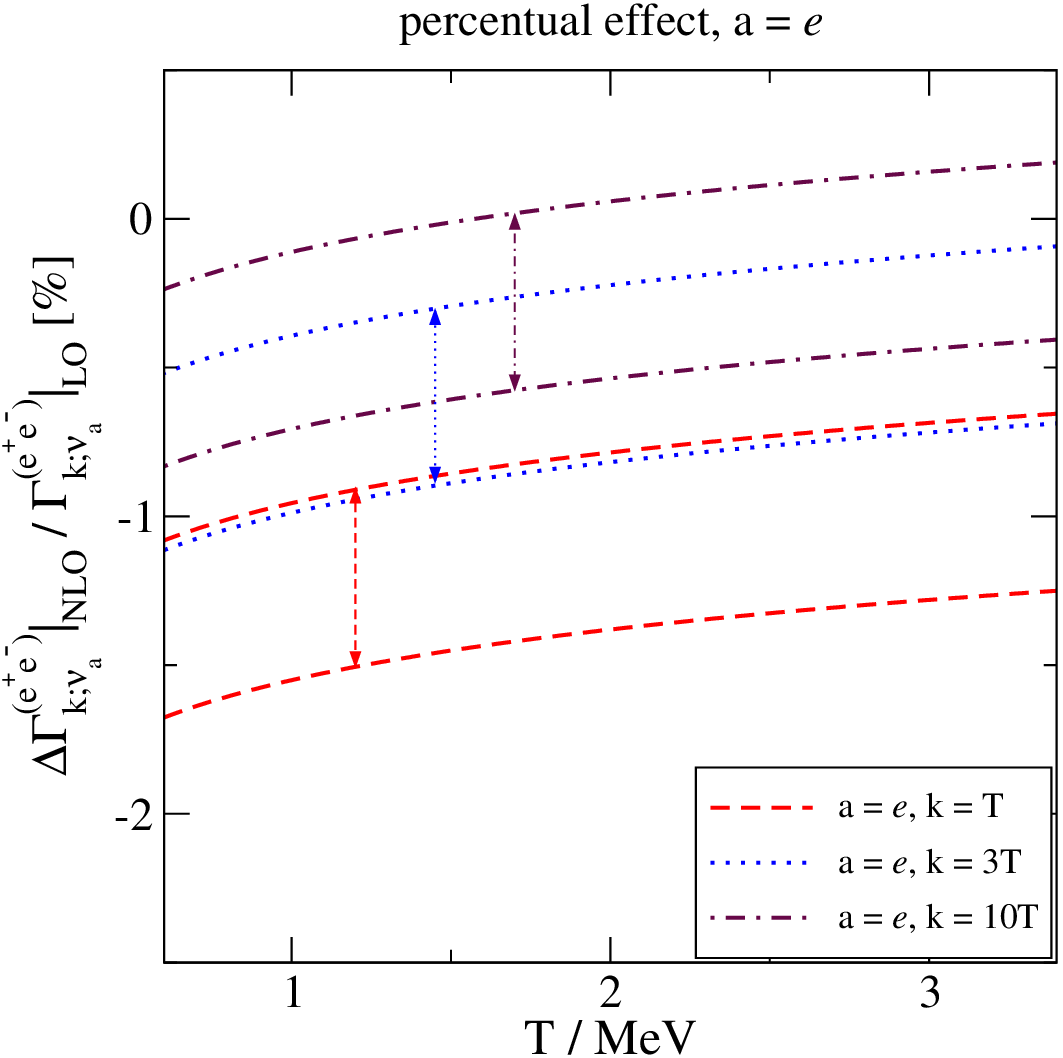}
  ~~~\epsfysize=7.0cm\epsfbox{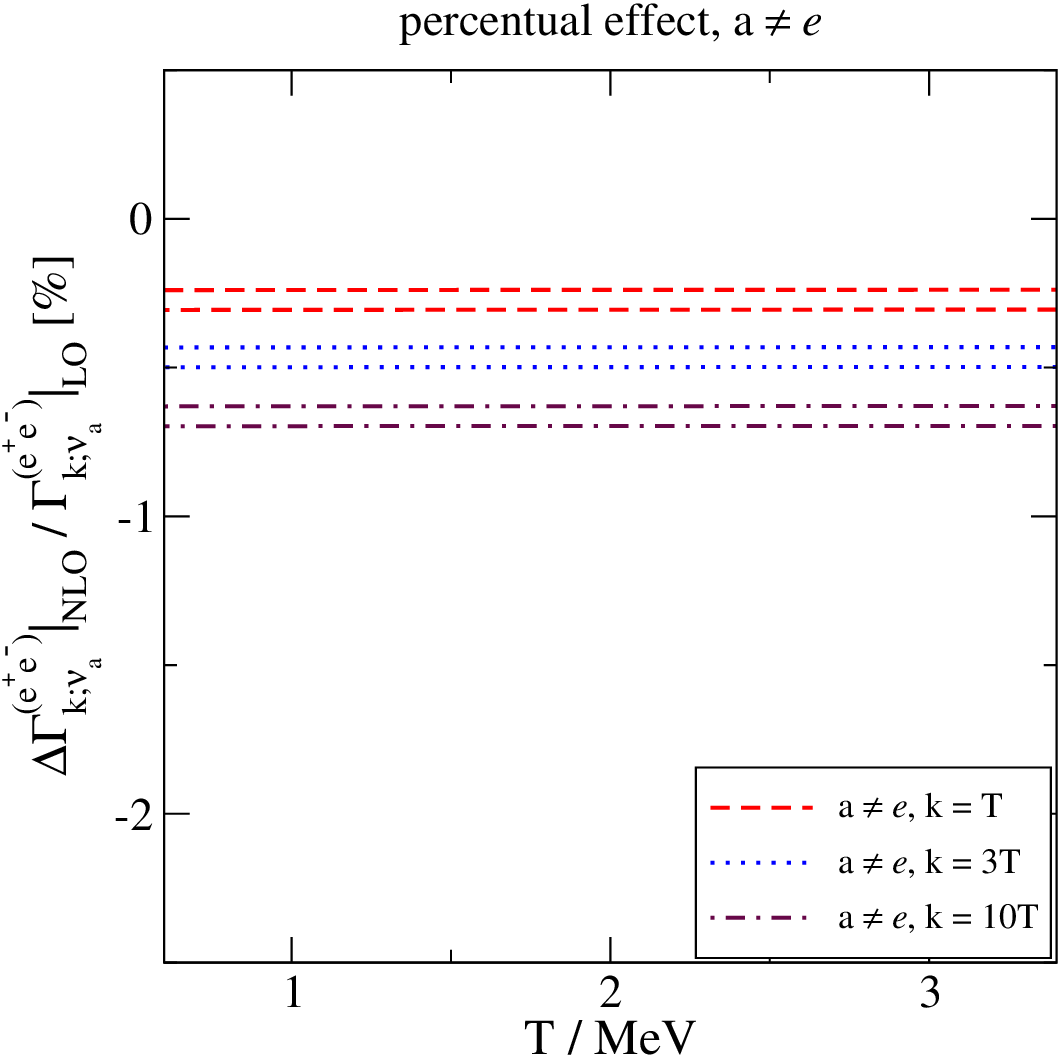}
}

\vspace*{1mm}

\caption[a]{\small 
 Numerical results for the percentual NLO QED contribution  
 to the $e^+_{ }e^-_{ }$ part of  
 the neutrino interaction rate, cf.\ \eq\nr{rho_8c}, 
 in massless QED, 
 with $\alpha^{ }_\rmi{em} = 1/137$.
 The uncertainty, indicated with a band, 
 reflects the low-energy hadronic contribution to the 
 coefficient $C^{ }_a$, as specified in \eq\nr{C_a}.
 Top row: 
 dependence on $k/T$ for fixed $T/$MeV, 
 for $a=e$ (left) and $a\neq e$ (right). 
 Bottom row: 
 dependence on $T$/MeV for fixed $k/T$, 
 for $a=e$ (left) and $a\neq e$ (right).
} 
\la{fig:results}
\end{figure}
%

%
\section{Conclusions}
\la{se:concl}

The general context of our investigation is 
the observable known as $N^{ }_\rmi{eff}$, 
parametrizing the energy density of the universe
at the time of primordial nucleosynthesis or photon decoupling 
through an effective number of massless neutrino species. 
In order to scrutinize
the current theoretical uncertainty of the Standard Model
prediction of $N^{ }_\rmi{eff}$, 
ref.~\cite{cemp} recently carried out an approximate NLO computation
of the energy transfer rate defined in \eq\nr{def_Q}.
They found a $\sim -5$\% QED correction to the energy transfer rate 
in the temperature range $T \in (1 ... 3)$\hspace*{0.3mm}MeV.
When this correction was inserted into momentum-averaged
kinetic equations describing neutrino decoupling~\cite{mea}, 
the fourth significant digit of the established Standard Model prediction
of~$N^{ }_\rmi{eff}$~\cite{Neffm2,Neffm1,Neff0} was  
asserted to decrease by one unit. 

The purpose of the present paper has been to compute NLO QED corrections
to the thermal interaction rate felt by 
active neutrinos in the same temperature range. 
The thermal interaction rate is defined by
\eq\nr{heli_proj} and influences neutrino physics 
via a momentum-dependent kinetic equation, 
of which \eq\nr{vn_1} is a simplified prototype. 
Therefore, it does {\em not} coincide
with the energy transfer rate in \eq\nr{def_Q}. 
However, it gets contributions
from the same diagrams (on our side, 
this is the subset called 
$s$-channel or inelastic processes, 
cf.\ \fig\ref{fig:processes}), 
and should reflect similar physics.  

In order to streamline the task, we have made 
a few simplifications concerning the 
overall statistical ensemble. In particular, we assumed 
the presence of a universal temperature, 
a vanishing electron mass, and vanishing lepton chemical potentials. 
However, we undertook no simplifications on the technical side, 
including in particular the full Bose and Fermi distributions, 
all processes depicted in \fig\ref{fig:processes},  
logarithmic running induced by the effective vertex in \eq\nr{ct_1}, 
and HTL resummation of soft $t$-channel photon exchange, 
which is otherwise IR divergent due to Bose enhancement. 
Moreover we have computed a differential
rate, displaying its momentum dependence. 

Our numerical result is shown in \fig\ref{fig:results}. 
We find a $\sim -(0...2)\%$ contribution, 
dominated by the $t$-channel diagrams in \fig\ref{fig:processes}(e,f,g), 
represented by the coefficients 
$B^{(t)}_{ }$, $C^{(t)}_{ }$ and $D^{(t)}_{ }$, 
illustrated numerically in appendix~C.
If we restrict to the $s$-channel processes, 
contributing to the observable in \eq\nr{def_Q}, 
the magnitude of the correction remains similar, 
but its sign is positive.
The overall magnitude of the $s$-channel 
correction is noticeably smaller than that found in ref.~\cite{cemp}, 
and its sign is opposite. 
It is well-known from other contexts that incomplete evaluations of NLO
corrections may overestimate a result, because cancellations
can go amiss (cf., e.g.,\ ref.~\cite{salvio}), however we are not
sure if this could be an explanation here. 
It might also be that the different observable of \eq\nr{def_Q}, 
and/or effects related to the electron mass,
could be responsible for the larger correction seen in ref.~\cite{cemp}. 
 
Another difference between our results and those in 
ref.~\cite{cemp} is that we observe a visible dependence 
of the relative NLO correction 
on the neutrino flavour $a$, whereas ref.~\cite{cemp}
stated the difference between the flavours to 
be less than 1\%. 
We remark that flavour dependence originates from 
the second and third lines of \eqs\nr{rho_8c} and \nr{ABCD},
corresponding to the processes in \fig\ref{fig:processes}(g), 
since the dependence on $a$ differs from that of the 
leading-order term, on the first line.
To be explicit, 
the $t$-channel contribution contains the diagram
\ba
  \nn[1mm]
  \hspace*{1.0cm}
  \extrNLOeX
  \qquad , \la{diag_V}
  \nn[-6mm]
\ea
where ``V'' indicates that the QED vertices are purely vectorlike.
As discussed around \eq\nr{gmuV}, the flavours $a\neq e$
are almost blind to the vectorlike coupling, and therefore
to this process. The same holds
for the $s$-channel reflections of this process, 
depicted as the first two diagrams
in \fig\ref{fig:processes}(g). 
At the same time, 
the flavour $a=e$ gets a visible contribution from the vectorlike coupling, 
in both the $t$ and $s$ channels, for instance through 
the logarithm $\sim \ln (T/m^{ }_e)$ in \eq\nr{ABCD}. 
Therefore, different flavours do feel different QED corrections. 

\vspace*{3mm}

Finally, we list possible directions of future research. 
On the mundane side, the effects of a finite 
electron mass could be incorporated at $\rmO(e^2)$, by making
use of the formalism developed in ref.~\cite{phasespace}, 
however this could turn out to be quite tedious numerically. 

A challenge of a more conceptual nature would be to generalize 
the framework so that non-equilibrium ensembles can be inserted for 
the different neutrino flavours. The QED plasma, 
inducing the NLO corrections
that we have computed here, would however maintain
the same form as in the present investigation. 

Last but not least, a significant role for the physics 
under consideration is played by the coefficient $C^{ }_a$, 
originating from the operator in \eq\nr{ct_1}. It would be 
important to have a good estimate of the magnitude
of its finite part, but
this is non-trivial, due to low-energy 
hadronic loops that affect $Z\gamma$ mixing. 
Our study made use of the range in \eq\nr{C_a}, 
extracted from refs.~\cite{eft2,eft3}, however the
error, visible as bands in \fig\ref{fig:results},  
could conceivably be reduced through further
theoretical and phenomenological work.  

%
\section*{Acknowledgements}

G.J.\ thanks Wouter Dekens for
a pointer to ref.~\cite{eft2}. 
M.L.\ thanks Miguel Escudero for discussions 
about the motivation for this work~\cite{cemp}, 
for drawing our attention to ref.~\cite{old}, 
and for comments on the manuscript, 
and Dietrich B\"odeker for many helpful discussions. 
G.J.\ was funded by the U.S.\ Department of Energy (DOE), 
under grant No.\ DE-FG02-00ER41132, 
and by the Agence Nationale de la Recherche (France), 
under grant ANR-22-CE31-0018 (AUTOTHERM).

%
\section*{Note added}

The very recent ref.~\cite{new} considered a subset of the
processes in our study, but included the electron mass.
Its conclusions are similar to ours, confirming in particular
that $m^{ }_e$ can generally be set to zero
at decoupling temperatures.

%
\appendix
\renewcommand{\thesection}{\Alph{section}}
\renewcommand{\thesubsection}{\Alph{section}.\arabic{subsection}}
\renewcommand{\theequation}{\Alph{section}.\arabic{equation}}

%
\section{Evaluation at leading order}

In this appendix we show how
the expression for the thermal neutrino
interaction rate can be simplified at leading order, 
by making use of symmetrizations of 
the sum-integration variables (cf.\ \se\ref{ss:lo}). 
We demonstrate how this simplification permits
for a numerical crosscheck of the full integral 
representation in \eq\nr{rho_qed} (cf.\ \se\ref{ss:numerics}).
Finally, we specify the real part of the correlator from 
\eq\nr{Vmunu}, as this is needed in \eq\nr{rho_8c}
(cf.\ \se\ref{ss:reV}). 

%
\subsection{Additional simplifications of the leading-order (LO) expression}
\la{ss:lo}

Let us return to \eqs\nr{Sigma_full_2}, \nr{Smunu_massless}
and \nr{heli_proj}, 
and consider the most basic scenario, 
namely assuming the absence of any chemical potentials,\footnote{%
 We have set the chemical potentials to vanish all along,
 but stress the point here, because it would otherwise 
 play an important role, 
 removing the symmetry between particles and antiparticles. 
 }
and setting the temperature of each
neutrino flavour equal. In this limit, 
the integrand develops additional symmetries, 
and consequently substitutions of sum-integration
variables allow us to put the result in a simple form. 

As a starting point, \eqs\nr{Sigma_full_2}, \nr{Smunu_massless} 
and \nr{heli_proj} yield (omitting contributions of $\rmO(e^2)$)
\ba
 && \hspace*{-1cm}
 \frac{ \Gamma^\rmii{LO}_{k;\nu^{ }_a} }{ {G_\rmii{F}^2}/({4k}) }
 \; = \; 
 \, 
 \Bigl\{ (2 {\textstyle \sum_b} ) + 
        \bigl[ \,
   \bigl( 2\delta^{ }_{a,e} - 1 + 4 \xW \bigr)^2_{ }
 + 1  \, \bigr] 
 \, \Bigr\} 
 \nn & \times & 
 \im \biggl\{ \, 
 \frac{
 \tr\bigl[\, 
 ( i \bsl{K} ) \gamma^{ }_\mu 
 (-i)( \bsl{K} + \bsl{P} ) \gamma^{ }_\nu (1 - \gamma^{ }_5 )
 \,\bigr]
 \tr\bigl[\, 
 (-i)( \bsl{P} + \bsl{Q} ) \gamma^{ }_\mu 
 (-i)( \bsl{Q} ) \gamma^{ }_\nu (1 - \gamma^{ }_5 )
 \,\bigr]
 }{ (K+P)^2_{ } (P+Q)^2_{ } Q^2_{ } }  
 \,\biggr\}
 \nn[3mm] 
 & + & 
 \, 
  (- 4 )
 \, 
 \im \biggl\{ \, 
 \frac{
 \tr \bigl[ \, 
 ( i \bsl{K} ) \gamma^{ }_\mu 
 (-i)( \bsl{K} + \bsl{P} ) \gamma^{ }_\nu
 (-i)( \bsl{P} + \bsl{Q} ) \gamma^{ }_\mu 
 (-i)( \bsl{Q} ) \gamma^{ }_\nu
 (1 - \gamma^{ }_5 )
 \, \bigr]
 }{ (K+P)^2_{ } (P+Q)^2_{ } Q^2_{ } } 
 \,\biggr\}
 \la{lo_line1}
 \;. \hspace*{5mm}
\ea
The first simplification originates from
the substitution $Q\to - P - Q$. This leaves
the denominator invariant, but symmetrizes the appearance of Feynman slashes
in the numerator. After some steps, this can be seen to guarantee 
that the $\gamma^{ }_5$-matrices drop out. 

Subsequently, 
taking the Dirac traces, 
we are left with various scalar
products of four-momenta. 
Some of these can be immediately completed
into squares, 
\ba
 K\cdot(K+P) & = & 
 \frac{1}{2} 
  \bigl[\,   
    \overbrace{K^2_{ }}^{\to 0} \; + \; 
    \overbrace{(K+P)^2_{ }}^{\rm no~cut} \; - \; 
    P^2_{ } 
  \,\bigr]
 \;, \la{squares_1} \\ 
 Q\cdot(P+Q) & = & 
 \frac{1}{2} 
  \bigl[\,   
    \overbrace{Q^2_{ }}^{\rm no~cut} \; + \;
    \overbrace{(P+Q)^2_{ }}^{\rm no~cut} \; - \; 
    P^2_{ } 
  \,\bigr]
 \;. \la{squares_2}
\ea
There are further scalar products which are not 
immediately completed into squares, however
they can also be brought into the desired form by permutations of momenta, 
\be
 \mathbbm{P}^{ }_1 \; \equiv \; \{ P\to -K + Q, \; Q\to - P - Q \}
 \;, \quad
 \mathbbm{P}^{ }_2 \; \equiv \; \{ P\to -K + Q, \; Q\to K + P \}
 \;. \la{perm}
\ee
Both leave the denominator invariant, 
whereas the ``non-canonical'' 
scalar products in the numerator get transferred as
\ba
 &&
 K\cdot(P+Q) \, Q\cdot (K+P)
  \; \stackrel{\mathbbm{P}^{ }_1\;}{\longrightarrow} \;
 K\cdot(K+P) \, (P+Q)\cdot Q 
 \;, \nn
 &&
 K\cdot Q \, (K+P)\cdot(P+Q)
  \; \stackrel{\mathbbm{P}^{ }_2\;}{\longrightarrow} \;
 K\cdot(K+P) \, Q\cdot (P+Q) 
 \;. 
\ea
Therefore we can again makes use of \eqs\nr{squares_1} and \nr{squares_2}. 
Implementing these steps and 
collecting the terms together, 
we find that \eq\nr{lo_line1} can be
reduced to 
\be
 \frac{ \Gamma^\rmii{LO}_{k;\nu^{ }_a} }{ {G_\rmii{F}^2}/({4k}) }
 \; = \; 
 \, 
 \Bigl\{ (2 {\textstyle \sum_b} )  + 
        \bigl[ \,
   \bigl( 2\delta^{ }_{a,e} - 1 + 4 \xW \bigr)^2_{ }
 + 1  \, \bigr] \,  + 2 
 \, \Bigr\} 
 \im \biggl\{ \, 
 \frac{ (-16) P^4
 }{ (K+P)^2_{ } (P+Q)^2_{ } Q^2_{ } }  
 \,\biggr\}
 \;. \la{lo_line2}
\ee

It remains to carry out the Matsubara sums. The outer sum-integral can 
be extracted from the $\beta$-coefficient in \eq\nr{rho_2}, yielding
\ba
 && \hspace*{-1cm} 
 \im \biggl\{ \, 
 \frac{ P^4
 }{ (K+P)^2_{ } (P+Q)^2_{ } Q^2_{ } }  
 \,\biggr\}
 \; = \;  
 \frac{1}{4\pi^2 k}
 \biggl(\,
 \overbrace{
  \int_{-\infty}^0 \! {\rm d} p^{ }_{-}  
  \int_0^k \! {\rm d} p^{ }_{+} 
 }^{\rm t-channel}
 -   
 \overbrace{
  \int_0^k \! {\rm d} p^{ }_{-} 
  \int_k^\infty \! {\rm d} p^{ }_+
  }^{\rm s-channel}
 \,\biggr)
 \nn & \times & 
 \, p\, \P^4_{ } \, 
  \bigl[\, 1 - \nF^{ }(k - p^0_{ }) + \nB^{ }(p^0_{ }) \,\bigr]
  \im\bigg\{\, 
    \frac{1}{  (P+Q)^2_{ } Q^2_{ }  }
  \,\biggr\}
 \;. \hspace*{4mm} \la{rho_5}
\ea
The inner sum-integral can be worked out in a similar fashion, giving
at first 
\ba
   \im\bigg\{\, 
    \frac{1}{  (P+Q)^2_{ } Q^2_{ }  }
  \,\biggr\}
 & = & 
 \frac{1}{16\pi p}
 \biggl\{\, 
 \overbrace{\theta(p^0_{ } - p)}^{\rm s-channel}
 \, 
 \int_{p^{ }_-}^{p^{ }_+} \! {\rm d}q \, 
 \bigl[\, 
   1 - \nF^{ }(q) - \nF^{ }(p^0_{ } - q)
 \,\bigr]
 \nn 
 & - & 
 2 \underbrace{\theta(p - |p^0_{ }|)}_{\rm t-channel}
 \, 
 \int_{p^{ }_+}^{\infty} \! {\rm d}q \, 
 \bigl[\, 
   \nF^{ }(q - p^0_{ }) - \nF^{ }(q) 
 \,\bigr]
 \,\biggr\}
 \;. \la{rho_5a}
\ea
Here we have made use of $\nF^{ }(-x)  = 1 - \nF^{ }(x)$
in order to represent the Fermi distributions with positive arguments. 
Subsequently the remaining integrals can be carried out in terms
of polylogarithms, 
\be
 \lnf(\omega) \; \equiv \; \ln \Bigl( 1 + e^{-\omega/T} \Bigr)
 \;, \quad\;
 \lif(\omega) \;\, \equiv \; \mbox{Li}^{ }_2 \Bigl(-e^{-\omega/T}\Bigr)
 \;, \quad\; 
 \ltf(\omega) \;\, \equiv \; \mbox{Li}^{ }_3 \Bigl(-e^{-\omega/T}\Bigr)
 \;, \la{polylogs}
\ee
with the latter two defined in anticipation of \eq\nr{rho00}. 
This converts \eq\nr{rho_5a} into
\ba
   \im\bigg\{\, 
    \frac{1}{  (P+Q)^2_{ } Q^2_{ }  }
  \,\biggr\}
 & = & 
 \overbrace{\theta(p^0_{ } - p)}^{\rm s-channel}
 \, 
 \frac{
     p + 
        2 T \, \bigl[ \,
                \lnf(p^{ }_{+}) - \lnf(p^{ }_{-})
             \, \bigr] }{16\pi p}
  \,
 \nn 
 & - & 
 2 \underbrace{\theta(p - |p^0_{ }|)}_{\rm t-channel}
 \, 
 \frac{ 
        T \, \bigl[ \,
                \lnf(-p^{ }_{-}) - \lnf(p^{ }_{+})
             \, \bigr] }{16\pi p}
  \,
 \;. \la{rho_5b}
\ea
Finally we can insert \eq\nr{rho_5b} into \eq\nr{rho_5}, 
obtaining a representation for 
$  \Gamma^\rmii{LO}_{k;\nu^{ }_a} $ similar to that found in the literature
(cf.,\ e.g.,\ ref.~\cite{broken} and references therein),
\ba
 \Gamma^\rmii{LO}_{k;\nu^{ }_a} 
 & = & 
 \Bigl\{ (2 {\textstyle \sum_b} ) +  
 \, \bigl[ \,
   \bigl( 2\delta^{ }_{a,e} - 1 + 4 \xW \bigr)^2_{ }
 + 1 
 \, \bigr] \, + 2  \,  
 \Bigr\} \,  
 \frac{G_\rmii{F}^2}{16\pi^3 k^2}
 \bigl[ \, 
  2 \Xi^t_{ }+ \Xi^s_{ }
 \, \bigr] 
 \;, \la{rho_6} \\[3mm]  
 \Xi^t_{ } & \equiv & 
  \int_{-\infty}^0 \! {\rm d} p^{ }_{-}  
  \int_0^k \! {\rm d} p^{ }_{+} 
 \, \P^4_{ }
 \, \bigl[\, 1 - \nF^{ }(k - p^0_{ }) + \nB^{ }(p^0_{ }) \,\bigr]
 \, \bigl\{ 
        T \, \bigl[ \,
                \lnf(-p^{ }_{-}) - \lnf(p^{ }_{+})
             \, \bigr]
  \,\bigr\} 
 \;, \nn \la{Xit} \\  
 \Xi^s_{ } & \equiv & 
  \int_0^k \! {\rm d} p^{ }_{-} 
  \int_k^\infty \! {\rm d} p^{ }_+
 \, \P^4_{ }
 \, \bigl[\, 1 - \nF^{ }(k - p^0_{ }) + \nB^{ }(p^0_{ }) \,\bigr]
 \, \bigl\{ 
     p + 
        2 T \, \bigl[ \,
                \lnf(p^{ }_{+}) - \lnf(p^{ }_{-})
             \, \bigr]
  \,\bigr\} 
 \;. \nn \la{Xis}
\ea
Here the prefactors within the curly brackets on the first line
of \eq\nr{rho_6} have been kept apart, in order to indicate their
origins in \eq\nr{Smunu_massless}. 

%
\subsection{Numerical crosscheck of the full integral representation}
\la{ss:numerics}

Let us compare the 
simplified representation in \eq\nr{rho_6} with the more complete
one following from \eq\nr{rho_qed}. For this we focus on the 
electron-positron contribution, which can be identified 
through the prefactor
$
 [ \,
   ( 2\delta^{ }_{a,e} - 1 + 4 \xW )^2_{ }
 + 1 
 \, ]
$
in \eq\nr{rho_6}. 

The leading-order electron-positron spectral functions, 
needed in \eq\nr{rho_qed},  
read~(the notation is from \eq\nr{polylogs})
\ba
 \rho_\rmii{V}^\rmii{LO} & \equiv & 
 - \bigl( 2 \rho^{ }_\rmii{T} + \rho^{ }_\rmii{L} \bigr)^\rmii{LO}_{ }
 \\[2mm]
  & \stackrel{p^{ }_+ > 0}{=} & 
  \frac{ \P^2_{ } }{4 \pi p }
  \Bigl\{
   p\, \theta(p^{ }_{-})  
 + 2 T \bigl[\, \lnf(p^{ }_{+}) - \lnf(|p^{ }_{-}|) \,\bigr]
  \Bigr\}
 \;, \la{rhoV} \\[3mm] 
 \rho_\rmii{00}^\rmii{LO} & \equiv & 
 - \frac{p^2}{\P^2_{ }}\,  
  \bigl( \rho^{ }_\rmii{L}  \bigr)^\rmii{LO}_{ }
 \\[2mm]
 & \stackrel{p^{ }_+ > 0}{=} & 
  \frac{ 1 }{12 \pi p }
  \Bigl\{ 
    p^3_{ }\theta(p^{ }_{-})
  + 12 p T^2 \bigl[\, \lif(p^{ }_{+})
        + \sign(p^{ }_{-})\, \lif(|p^{ }_{-}|) \,\bigr]
  \nn & + & 
  \, 24 T^3 \bigl[\, \ltf(p^{ }_{+}) - \ltf(| p^{ }_{-} |) \,\bigr]
  \Bigr\} 
 \;. \la{rho00}
\ea
The limits at $p,|p^0_{ }| \ll \pi T$ can be found 
in \eqs\nr{rhoTir} and \nr{rhoLir}. 
The values at $p^{ }_+ < 0$ could be obtained 
by making use of the antisymmetry
in $p^{ }_0 \to -p^{ }_0$, however they are not needed in practice.
Inserting these, \eqs\nr{rho_4}, \nr{e_part} and \nr{rho_qed} become
\ba
 \Gamma^\rmii{LO}_{k;\nu^{ }_a}
 & \supset & 
 - 
 \, \bigl[ \,
   \bigl( 2\delta^{ }_{a,e} - 1 + 4 \xW \bigr)^2_{ }
 + 1 
 \, \bigr] \,
 \frac{G_\rmii{F}^2}{8\pi^2k^2} 
 \biggl(\,
  \int_{-\infty}^0 \! {\rm d} p^{ }_{-}  
  \int_0^k \! {\rm d} p^{ }_{+} 
 -   
  \int_0^k \! {\rm d} p^{ }_{-} 
  \int_k^\infty \! {\rm d} p^{ }_+
 \,\biggr)
 \nn & \times & 
 \, 
  \frac{p\, \P^2_{ }}{2}
 \, 
  \bigl[\, 1 - \nF^{ }(k - p^0_{ }) + \nB^{ }(p^0_{ }) \,\bigr]
 \nn[2mm] 
  & \times & 
 \biggl\{
   \rho^\rmii{LO}_\rmii{V}  
   \biggl[
     1 + \biggl( \frac{2k - p^0_{ }}{p} \biggr)^2_{ } 
   \biggr]
   + 
   \frac{\rho^\rmii{LO}_\rmii{00} \, \P^2_{ }}{p^2} \,
   \biggl[
     1  - 3\, \biggl( \frac{2k - p^0_{ }}{p} \biggr)^2_{ } 
   \biggr]
 \biggr\}
 \;. \la{rho_7}
\ea

Even if \eqs\nr{rho_6} and \nr{rho_7} look different, with the latter
containing a much more complicated integrand, both can be evaluated 
numerically, confirming their equivalence. 

%
\subsection{Real part of the vector correlator}
\la{ss:reV}

Let us decompose the real part of the vector correlator from 
\eq\nr{Vmunu} like in \eq\nr{decomposition}, 
\be
 \re V^{\mu\nu}_{\P} = 
 \mathbbm{P}^{\mu\nu}_\rmii{T} \, \chi^{ }_\rmii{T} + 
 \mathbbm{P}^{\mu\nu}_\rmii{L} \, \chi^{ }_\rmii{L}
 \;. \la{chi_def}
\ee
The results have divergent vacuum parts and finite thermal parts. 
Writing $D = 4 - 2\epsilon$ and denoting the 
scale parameter of the $\msbar$ scheme by
$\bmu^2 \equiv 4\pi \mu^2 e^{-\gammaE}$, the vacuum parts read
\be
 \chi^\rmii{LO}_\rmii{T} \;\bigr|^{ }_\rmi{vac}
 \; = \;  
 \chi^\rmii{LO}_\rmii{L} \;\bigr|^{ }_\rmi{vac}
 \; = \; 
 - \frac{4\P^2_{ }}{3}
   \frac{\mu^{-2\epsilon}_{ }}{(4\pi)^2}
  \biggl(
    \frac{1}{\epsilon} + \ln\frac{\bmu^2_{ }}{\P^2_{ }} + \fr53
  + 
 \rmO(\epsilon)
  \biggr)
 \;. \la{chi_vac}
\ee
As for the thermal parts, they are best represented through the 
same linear combinations as in \eqs\nr{rhoV} and \nr{rho00}. 
For these we find
\ba
 \chi^\rmii{LO}_\rmii{V} \bigr|^{ }_\T & \equiv & 
 - \bigl( 2 \chi^{ }_\rmii{T} + \chi^{ }_\rmii{L} \bigr)^\rmii{LO}_{ }
 \bigr|^{ }_\T
 \\[2mm]
 & = & 
 4 \int_q
 \frac{\nF^{ }(q)}{q}
 \biggl\{\;
  -2 - \frac{\P^2_{ }}{4 p q}
 \ln \biggl|
       \frac{ 1 - [(p+2q)/p^0_{ }]^2 }
            { 1 - [(p-2q)/p^0_{ }]^2 } 
     \biggr| 
 \;\biggr\}
 \;, \la{chiV} \\[3mm] 
 \chi^\rmii{LO}_\rmii{00} \bigr|^{ }_\T & \equiv & 
 - \frac{p^2}{\P^2_{ }}\,  
  \bigl( \chi^{ }_\rmii{L}  \bigr)^\rmii{LO}_{ }
  \bigr|^{ }_\T
  \\[2mm]
 & = & 
 4 \int_q
 \frac{\nF^{ }(q)}{q}
 \biggl\{\;
  1  + \frac{\P^2_{ } + 4 q^2}{8 p q}
 \ln \biggl|
       \frac{ 1 - [(p+2q)/p^0_{ }]^2 }
            { 1 - [(p-2q)/p^0_{ }]^2 } 
     \biggr| 
  - \frac{p^0_{ }}{2p}
 \ln \biggl|
       \frac{ 1 - [2q/(p-p^0_{ })]^2 }
            { 1 - [2q/(p+p^0_{ })]^2 } 
     \biggr| 
 \;\biggr\}
 \;, \nn \la{chi00}
\ea
where 
$
 \int_q \equiv \int_0^\infty \! {\rm d}q\, q^2 / (2\pi^2)
$.
The limits at $p,|p^0_{ }| \ll \pi T$ can be found 
in \eqs\nr{chiTir} and \nr{chiLir}. 

%
\section{Leading-logarithmic part of the NLO correction}
\la{se:log}

We show here how the coefficient 
of the logarithm in \eq\nr{Dt_log},
$D^{(t)}_1$,
can be derived analytically, 
starting from the HTL-resummed part of \eq\nr{rho_8c}.
Physically, this corresponds to an interference term between 
the amplitude in \eq\nr{diag_V} 
and a tree-level $t$-channel process.

Parametrizing the interaction rate according to \eq\nr{ABCD}, and 
considering the contribution from the domain of small spatial momenta, 
$0 < p < p^{ }_*$, where $e T \ll p^{ }_* \ll \{k,\pi T\}$ is 
an intermediate cutoff, we can write
\ba
 D^{(t)}_{ } & \supset & 
 \frac{1}{k^3_{ }T^4_{ }}
 \overbrace{ 
  \frac{1}{2} \int_0^{p^{ }_*} \! {\rm d}p 
  \int_{-p}^{+p} \! {\rm d} p^0_{ }
 }^{\supset\, 
  \int_{-\infty}^0 \! {\rm d} p^{ }_{-}  
  \int_0^k \! {\rm d} p^{ }_{+}
   }
 \, 2 \, {p}\, \P^2_{ } 
 \overbrace{
             \frac{T}{p^0_{ }}
           }^{ \supset\,
               n_\rmiii{B}^{ }(p^0_{ })
             }
   \bigl(
    \rho^\rmii{IR}_\rmii{T} \chi^\rmii{IR}_\rmii{T} \mathcal{R}^*_\rmii{T}
  - \rho^\rmii{IR}_\rmii{L} \chi^\rmii{IR}_\rmii{L} \mathcal{R}^*_\rmii{L}
   \bigr)  
 \overbrace{ 
 \frac{4k^2_{ }}{p^2}
           }^{ \supset\, 
               \bigl( \frac{2k - p^0_{ }}{p} \bigr)^2_{ }
             }
 \nn[2mm] 
 & \stackrel{p^0_{ } = x\,p}{=} & 
 \frac{4}{k T^3}
 \int_0^{p^{ }_*}\! {\rm d}p \, p 
 \int_{-1}^{+1} \! \frac{{\rm d}x \, (x^2_{ }-1)}{x}
   \, 
   \bigl(
    \rho^\rmii{IR}_\rmii{T} \chi^\rmii{IR}_\rmii{T} \mathcal{R}^*_\rmii{T}
  - \rho^\rmii{IR}_\rmii{L} \chi^\rmii{IR}_\rmii{L} \mathcal{R}^*_\rmii{L}
   \bigr)^{ }_{p^0_{ } = x\, p }
 \;.   
\ea 
Here, for $n\in\{\rmi{T,L}\}$, \eq\nr{R} can be expressed as 
\be
 \mathcal{R}^{*}_n = 
 \re\biggl\{\, \frac{1}{p^2_{ }(x^2_{ } - 1)
  - e^2_{ } (\chi^\rmii{IR}_n + i \rho^\rmii{IR}_n)} 
  \,\biggr\} 
 =
 \re\biggl\{\,
    \frac{1}{p^2_{ } - \Pi^{ }_n} 
    \,\biggr\}
    \,\frac{1}{x^2_{ } - 1}
 \;, 
\ee 
where we have defined 
\be
 \Pi^{ }_n \; \equiv \; \frac{e^2_{ } 
 (\chi^\rmii{IR}_n + i \rho^\rmii{IR}_n)}{x^2_{ } - 1}
 \;. \la{Pin_def} 
\ee
We note that after the substitution $p^0_{ }= x\, p$, the self-energies
$\chi^\rmii{IR}_n$ and $\rho^\rmii{IR}_n$ 
from \eqs\nr{rhoTir}--\nr{chiLir}, 
and consequently $\Pi^{ }_n$ from \eq\nr{Pin_def}, 
are functions only of $x$.
Therefore, we can carry out the integral over $p$, 
\be
 \int_0^{p^{ }_*} \! \frac{ {\rm d}p \, p}{p^2_{ } - \Pi^{ }_n}
 = 
 \frac{1}{2} \ln \biggl( \frac{p_*^2 - \Pi^{ }_n}{-\Pi^{ }_n} \biggr)
 = 
 \frac{1}{2} 
 \biggl[
    \ln\biggl( \frac{p_*^2}{e^2_{ } T^2_{ }} \biggr) 
  + \ln\biggl( \frac{e^2_{ } T^2_{ }}{-\Pi^{ }_n} \biggr)
  + \rmO\biggl( \frac{-\Pi^{ }_n}{p_*^2}\biggr)
 \biggr]
 \;. 
\ee
The first term yields
\ba
 D^{(t)}_{ } & \supset & 
 \frac{2}{ k T^3_{ } }\,
 \ln\biggl( \frac{p_*^2}{e^2_{ } T^2_{ }} \biggr)\,
 \int_{-1}^{+1} \! \frac{{\rm d}x}{x}
   \,
   \bigl(
    \rho^\rmii{IR}_\rmii{T} \chi^\rmii{IR}_\rmii{T} 
  - \rho^\rmii{IR}_\rmii{L} \chi^\rmii{IR}_\rmii{L} 
   \bigr)^{ }_{p^0_{ } = x\, p } 
 \nn[2mm]
 & \stackrel{\rmii{\nr{rhoTir}--\nr{chiLir}}}{=} & 
 \frac{\pi T}{ 36 k }\,
 \ln\biggl( \frac{p_*^2}{e^2_{ } T^2_{ }} \biggr)\,
 \int_{-1}^{+1} \! {\rm d}x \, (x^2_{ }  - 1)\,
 \biggl[\,
   4 - 3x^2_{ }+ \frac{3x(x^2_{ } - 1)}{2}\ln\frac{1+x}{1-x}
 \,\biggr]
 \nn[2mm]
 & = & 
 - \frac{\pi T}{ 9 k }\,
 \ln\biggl( \frac{p_*^2}{e^2_{ } T^2_{ }} \biggr)\,
 \;. \la{Dt_log_derivation}
\ea
The coefficient of $\ln(1/e^2_{ })$ yields $D^{(t)}_1$ 
from \eq\nr{Dt_log}.

%
\section{Numerical results for coefficient functions}
\la{se:plots}

Our main results are captured by the coefficients 
$A$, $B$, $C$ and $D$, defined in \eq\nr{ABCD}.
Examples of their numerical values are displayed 
in table~\ref{table:coeffs}, permitting for a precise 
evaluation of the full result at selected momenta.  
The overall pattern may be easier to appreciate from a plot, 
whence we show one in \fig\ref{fig:plots}.  

%
\begin{table}[t]

\small{
\begin{center}
\begin{tabular}{cccccccccc} 
 $\!\! k / T  \!\!\!$ & 
 $ A^{(s)}_{ }$ & 
 $ A^{(t)}_{ }$ & 
 $ B^{(s)}_{ }$ & 
 $ B^{(t)}_{ }$ & 
 $ C^{(s)}_{ }$ & 
 $ C^{(t)}_{ }$ & 
 $ D^{(s)}_{ }$ & 
 $ D^{(t)}_{1}$ & 
 $ D^{(t)}_{2}$   
 \\[2mm]
 \hline
 \\[-4mm] 
 0.5  & 
 1.80721  & 
 7.31870  & 
 0.142  & 
-0.192  & 
 3.120  & 
 4.132  & 
 0.098  & 
-0.698   & 
-0.654   
   \\[0.5mm]  
 1.0  & 
 1.66528  & 
 6.70427  & 
 0.061  & 
-0.308  & 
 4.013  & 
 8.224  & 
 0.037  & 
-0.349   & 
-0.495   
   \\[0.5mm]  
 2.0  & 
 1.53556  & 
 6.14451  & 
 0.042  & 
-0.372  & 
 4.741  & 
 11.56  & 
 0.011  & 
-0.175   & 
-0.319   
   \\[0.5mm]  
 3.0  & 
 1.50561  & 
 6.01450  & 
 0.037  & 
-0.419  & 
 5.244  & 
 13.66  & 
 0.005  & 
-0.116   & 
-0.238   
   \\[0.5mm]  
 4.0  & 
 1.50956  & 
 6.03093  & 
 0.035  & 
-0.459  & 
 5.681  & 
 15.38  & 
 0.003  & 
-0.087   & 
-0.192   
   \\[0.5mm]  
 5.0  & 
 1.52175  & 
 6.08366  & 
 0.034  & 
-0.493  & 
 6.059  & 
 16.86  & 
 0.002  & 
-0.070   & 
-0.160   
   \\[0.5mm]  
 6.0  & 
 1.53387  & 
 6.13650  & 
 0.033  & 
-0.522  & 
 6.381  & 
 18.11  & 
 0.002  & 
-0.058   & 
-0.138   
   \\[0.5mm]  
 7.0  & 
 1.54392  & 
 6.18043  & 
 0.032  & 
-0.545  & 
 6.656  & 
 19.19  & 
 0.001  & 
-0.050   & 
-0.121   
   \\[0.5mm]  
 8.0  & 
 1.55185  & 
 6.21507  & 
 0.032  & 
-0.565  & 
 6.894  & 
 20.12  & 
 0.001  & 
-0.044   & 
-0.108   
   \\[0.5mm]  
 9.0  & 
 1.55808  & 
 6.24214  & 
 0.032  & 
-0.583  & 
 7.103  & 
 20.94  & 
 0.001  & 
-0.039   & 
-0.098   
   \\[0.5mm]  
 10.0  & 
 1.56303  & 
 6.26349  & 
 0.031  & 
-0.598  & 
 7.288  & 
 21.67  & 
 0.001  & 
-0.035   & 
-0.089   
   \\[0.5mm]  
 15.0  & 
 1.57756  & 
 6.32427  & 
 0.031  & 
-0.653  & 
 7.988  & 
 24.42  & 
 0.001  & 
-0.023   & 
-0.062   
   \\[0.5mm]  
 \hline 
\end{tabular} 
\end{center}
}


\caption[a]{\small
 The coefficients from \eq\nr{ABCD}
 and \nr{Dt_log}, as a function of the momentum $k/T$.
 The same results are represented as plotted curves
 in \fig\ref{fig:plots}.
 }
\label{table:coeffs}
\end{table}
%

%
\begin{figure}[t]

\centerline{
     \epsfysize=7.0cm\epsfbox{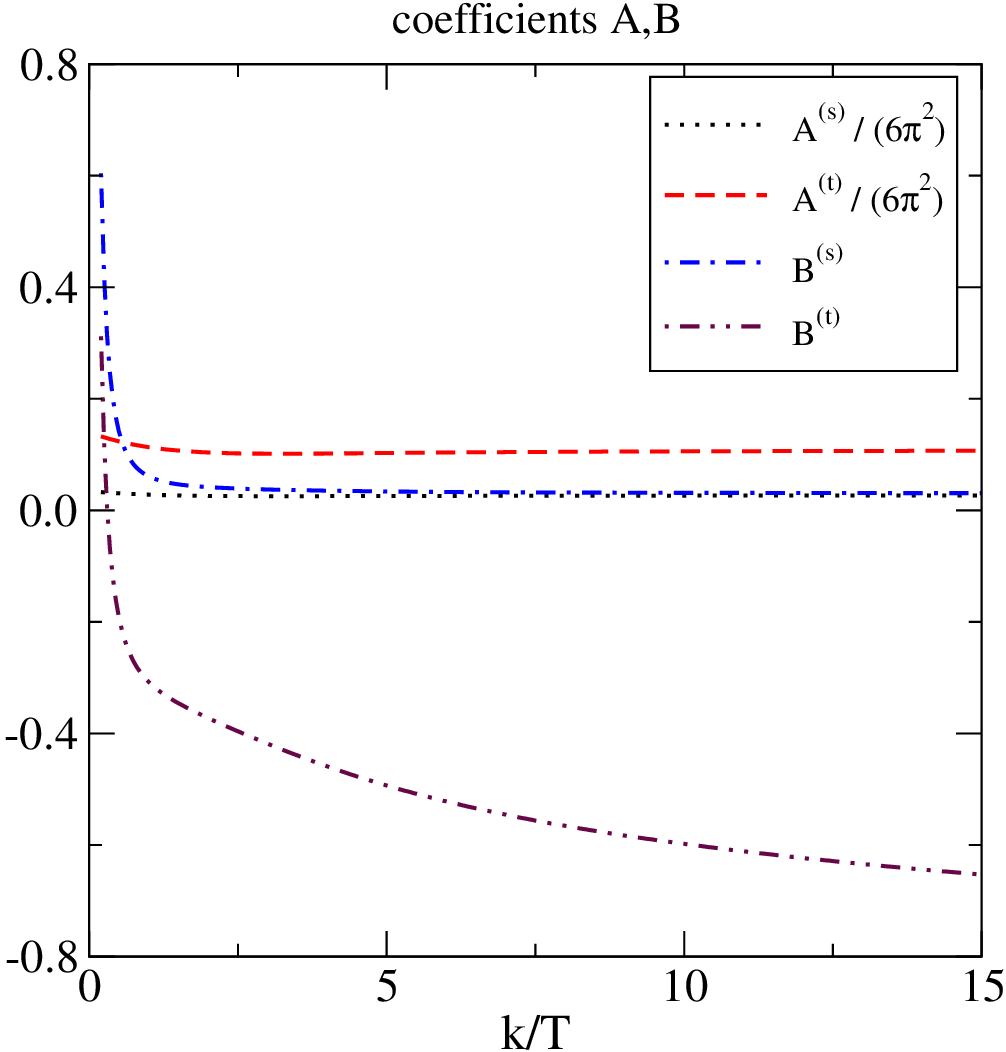}
  ~~~\epsfysize=7.0cm\epsfbox{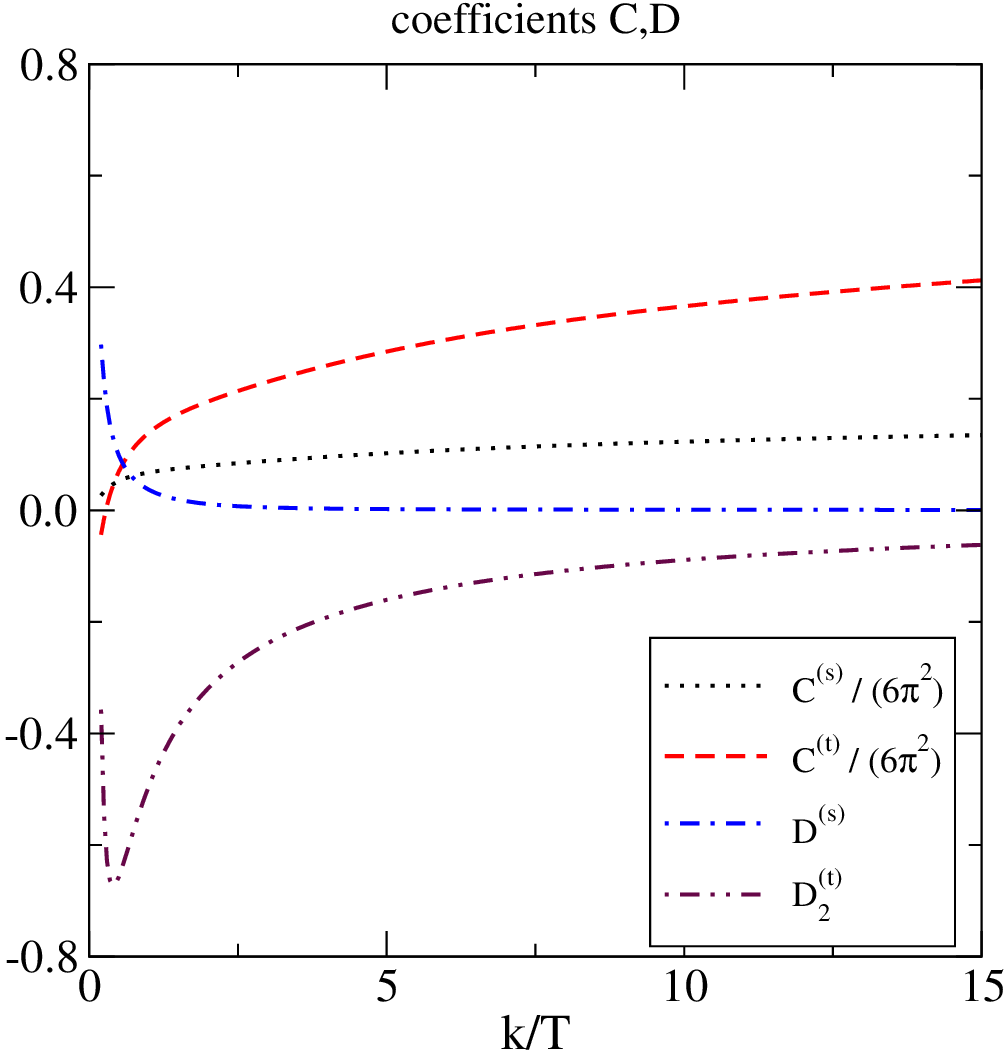}
}

\vspace*{1mm}

\caption[a]{\small 
 The coefficients from \eqs\nr{ABCD}
 and \nr{Dt_log}, as a function of the momentum $k/T$.
 The precise values at a number of selected momenta
 can be found in table~\ref{table:coeffs}, and the 
 coefficient $D^{(t)}_1$ is given analytically in \eq\nr{Dt_log}.
 The coefficients $A$ and $C$ have been multiplied by 
 $1/(6\pi^2_{ })\approx 0.0169$, so that their 
 magnitudes can be meaningfully compared with those of $B$ and $D$, 
 cf.\ \eq\nr{ABCD}. 
} 
\la{fig:plots}
\end{figure}
%

\small{
%

}

\end{document}